\documentclass[numberedappendix,letterpaper]{emulateapj}
\newcommand{\deltab}{deluxetable*}

\usepackage{apjfonts}
\usepackage{natbib}
\usepackage{amsmath}
\usepackage{color,colordvi}
\usepackage[mediumspace,Gray,squaren]{SIunits}
\usepackage{hyperref}

\definecolor{darkblue}{rgb}{0,0,0.3}
\definecolor{darkgreen}{rgb}{0,0.3,0}
\hypersetup{
  pdftitle = {ACT: Beam Profiles and First SZ Clusters},
  pdfauthor = {Hincks et al.},
  colorlinks,%
  citecolor=black,%
  filecolor=black,%
  linkcolor=blue,%
  urlcolor=red
}

\newcommand{\arone}{148\,GHz}
\newcommand{\artwo}{218\,GHz}
\newcommand{\arthree}{277\,GHz}

\newcommand{\commentx}[1]{}

\newcommand{\nsr}    {\ensuremath{\mathrm{\nano\steradian}}} 

\providecommand{\e}[1]{\ensuremath{\!\!\times\!\!10^{#1}}} 

\newcommand{\ra}[3]   
   {\makebox[1.5em][r]{#1}\makebox[1.5em][r]{#2} \makebox[2em][r]{#3}}

\newcommand{\hms}[3]  
   {${#1}^{\mathrm{h}}{#2}^{\mathrm{m}}{#3}^{\mathrm{s}}$}

\newcommand{\hmin}[2]  
   {\ensuremath{{#1}^{\mathrm{h}}{#2}^{\mathrm{m}}}}
   
\newcommand{\hours}[1]  
   {\ensuremath{{#1}^{\mathrm{h}}}}

\newcommand{\dms}[3]  
   {\ensuremath{{#1}\degree{#2}\arcminute{#3}\arcsecond}}

\newcommand{\dm}[2]  
   {\ensuremath{{#1}\degree{#2}\arcminute}}

\newcommand{\ukcmb}  
           {\ensuremath{\micro \kelvin_\mathrm{cmb}}}

\newcommand{\uk}  
           {\ensuremath{\micro \kelvin}}

\newcommand{\fdeg} 
           {\hbox{$.\!\!^{\circ}$}}

\usepackage{color}

\renewcommand{\bm}[1]{\textbf{#1}}
\newcommand{\bvec}[1]{{#1}}
\newcommand{\bh}[1]{\hat{\bvec{#1}}}
\newcommand{\wt}[1]{\widetilde{#1}}
\newcommand{\imat}{\bm{1}}
\newcommand{\dytwo}{0.2} 
\newcommand{\dyfour}{0.6} 
\newcommand{\dysix}{1.1} 
\newcommand{\bs}[1]{\boldsymbol{#1}}

\shorttitle{ACT: BEAM PROFILES AND FIRST SZ CLUSTERS}
\shortauthors{HINCKS ET AL.}

\begin{document}

\title{The Atacama Cosmology Telescope (ACT):  Beam Profiles and
       First SZ Cluster Maps}

\journalinfo{ApJS, 191:423--438, 2010 December}
\submitted{Submitted 2009 June 19; accepted 2010 November 3;
           published 2010 December 2}

\begin{abstract}
  The Atacama Cosmology Telescope (ACT) is currently observing the cosmic
  microwave background with arcminute resolution at \arone, \artwo,
  and \arthree.    In this paper, we present ACT's first results.
  Data have been analyzed using a maximum-likelihood
  map-making method which uses B-splines to model and remove the
  atmospheric signal.  It has been used to make
  high-precision beam maps from which we determine the experiment's window
  functions.  This beam information directly impacts all subsequent
  analyses of the data.  We also used the method to map a sample of galaxy
  clusters via the Sunyaev-Zel'dovich (SZ) effect, and show five
  clusters previously detected with X-ray or SZ observations.  We
  provide integrated Compton-$y$ measurements for each cluster.  Of
  particular interest is our detection of  the $z = 0.44$ component of
  A3128 and our current non-detection of the low-redshift part,
  providing strong evidence that the further cluster is more massive
  as suggested by X-ray measurements.  This is a compelling example of
  the redshift-independent mass selection of the SZ effect. 
\end{abstract}

\author{
A.~D.~Hincks\altaffilmark{1},
V.~Acquaviva\altaffilmark{2,3},
P.~A.~R.~Ade\altaffilmark{4},
P.~Aguirre\altaffilmark{5},
M.~Amiri\altaffilmark{6},
J.~W.~Appel\altaffilmark{1},
L.~F.~Barrientos\altaffilmark{5},
E.~S.~Battistelli\altaffilmark{7,6},
J.~R.~Bond\altaffilmark{8},
B.~Brown\altaffilmark{9},
B.~Burger\altaffilmark{6},
J.~Chervenak\altaffilmark{10},
S.~Das\altaffilmark{11,1,3},
M.~J.~Devlin\altaffilmark{12},
S.~R.~Dicker\altaffilmark{12},
W.~B.~Doriese\altaffilmark{13},
J.~Dunkley\altaffilmark{14,1,3},
R.~D\"unner\altaffilmark{5},
T.~Essinger-Hileman\altaffilmark{1},
R.~P.~Fisher\altaffilmark{1},
J.~W.~Fowler\altaffilmark{1},
A.~Hajian\altaffilmark{8,3,1},
M.~Halpern\altaffilmark{6},
M.~Hasselfield\altaffilmark{6},
C.~Hern\'andez-Monteagudo\altaffilmark{15},
G.~C.~Hilton\altaffilmark{13},
M.~Hilton\altaffilmark{16,17},
R.~Hlozek\altaffilmark{14},
K.~M.~Huffenberger\altaffilmark{18},
D.~H.~Hughes\altaffilmark{19},
J.~P.~Hughes\altaffilmark{2},
L.~Infante\altaffilmark{5},
K.~D.~Irwin\altaffilmark{13},
R.~Jimenez\altaffilmark{20},
J.~B.~Juin\altaffilmark{5},
M.~Kaul\altaffilmark{12},
J.~Klein\altaffilmark{12},
A.~Kosowsky\altaffilmark{9},
J.~M.~Lau\altaffilmark{21,22,1},
M.~Limon\altaffilmark{23,12,1},
Y.-T.~Lin\altaffilmark{24,3,5},
R.~H.~Lupton\altaffilmark{3},
T.~A.~Marriage\altaffilmark{25,3},
D.~Marsden\altaffilmark{12},
K.~Martocci\altaffilmark{26,1},
P.~Mauskopf\altaffilmark{4},
F.~Menanteau\altaffilmark{2},
K.~Moodley\altaffilmark{16,17},
H.~Moseley\altaffilmark{10},
C.~B.~Netterfield\altaffilmark{27},
M.~D.~Niemack\altaffilmark{13,1},
M.~R.~Nolta\altaffilmark{8},
L.~A.~Page\altaffilmark{1},
L.~Parker\altaffilmark{1},
B.~Partridge\altaffilmark{28},
H.~Quintana\altaffilmark{5},
B.~Reid\altaffilmark{20,1},
N.~Sehgal\altaffilmark{21},
J.~Sievers\altaffilmark{8},
D.~N.~Spergel\altaffilmark{3},
S.~T.~Staggs\altaffilmark{1},
O.~Stryzak\altaffilmark{1},
D.~S.~Swetz\altaffilmark{12,13},
E.~R.~Switzer\altaffilmark{26,1},
R.~Thornton\altaffilmark{12,29},
H.~Trac\altaffilmark{30,3},
C.~Tucker\altaffilmark{4},
L.~Verde\altaffilmark{20},
R.~Warne\altaffilmark{16},
G.~Wilson\altaffilmark{31},
E.~Wollack\altaffilmark{10},
 and
Y.~Zhao\altaffilmark{1}
}
\affiliation{$^{1}$ Joseph Henry Laboratories of Physics, Jadwin Hall,
Princeton University, Princeton, NJ 08544, USA}
\affiliation{$^{2}$ Department of Physics and Astronomy, Rutgers,
The State University of New Jersey, Piscataway, NJ 08854-8019, USA}
\affiliation{$^{3}$ Department of Astrophysical Sciences, Peyton Hall,
Princeton University, Princeton, NJ 08544, USA}
\affiliation{$^{4}$ School of Physics and Astronomy, Cardiff University, The Parade, Cardiff, Wales CF24 3AA, UK}
\affiliation{$^{5}$ Departamento de Astronom{\'{i}}a y Astrof{\'{i}}sica,
Facultad de F{\'{i}}sica, Pontific\'{i}a Universidad Cat\'{o}lica de Chile,
Casilla 306, Santiago 22, Chile}
\affiliation{$^{6}$ Department of Physics and Astronomy, University of
British Columbia, Vancouver, BC V6T 1Z4, Canada}
\affiliation{$^{7}$ Department of Physics, University of Rome ``La Sapienza'',
Piazzale Aldo Moro 5, I-00185 Rome, Italy}
\affiliation{$^{8}$ Canadian Institute for Theoretical Astrophysics, University of
Toronto, Toronto, ON M5S 3H8, Canada}
\affiliation{$^{9}$ Department of Physics and Astronomy, University of Pittsburgh,
Pittsburgh, PA 15260, USA}
\affiliation{$^{10}$ Code 553/665, NASA/Goddard Space Flight Center,
Greenbelt, MD 20771, USA}
\affiliation{$^{11}$ Berkeley Center for Cosmological Physics, LBL and
Department of Physics, University of California, Berkeley, CA 94720, USA}
\affiliation{$^{12}$ Department of Physics and Astronomy, University of
Pennsylvania, 209 South 33rd Street, Philadelphia, PA 19104, USA}
\affiliation{$^{13}$ NIST Quantum Devices Group, 325
Broadway Mailcode 817.03, Boulder, CO 80305, USA}
\affiliation{$^{14}$ Department of Astrophysics, Oxford University, Oxford OX1 3RH,
UK}
\affiliation{$^{15}$ Max Planck Institut f\"ur Astrophysik, Postfach 1317,
D-85741 Garching bei M\"unchen, Germany}
\affiliation{$^{16}$ Astrophysics and Cosmology Research Unit, School of
Mathematical Sciences, University of KwaZulu-Natal, Durban, 4041,
South Africa}
\affiliation{$^{17}$ Centre for High Performance Computing, CSIR Campus, 15 Lower
Hope Street, Rosebank, Cape Town, South Africa}
\affiliation{$^{18}$ Department of Physics, University of Miami, Coral Gables,
FL 33124, USA}
\affiliation{$^{19}$ Instituto Nacional de Astrof\'isica, \'Optica y
Electr\'onica (INAOE), Tonantzintla, Puebla, Mexico}
\affiliation{$^{20}$ ICREA \& Institut de Ciencies del Cosmos (ICC), University of
Barcelona, Barcelona 08028, Spain}
\affiliation{$^{21}$ Kavli Institute for Particle Astrophysics and Cosmology, Stanford
University, Stanford, CA 94305-4085, USA}
\affiliation{$^{22}$ Department of Physics, Stanford University, Stanford, CA 94305-4085,
USA}
\affiliation{$^{23}$ Columbia Astrophysics Laboratory, 550 West 120th Street, Mail Code
5247, New York, NY 10027, USA}
\affiliation{$^{24}$ Institute for the Physics and Mathematics of the Universe,
The University of Tokyo, Kashiwa, Chiba 277-8568, Japan}
\affiliation{$^{25}$ Department of Physics and Astronomy, The Johns Hopkins University, 3400
North Charles Street, Baltimore 21218-2686, MD}
\affiliation{$^{26}$ Kavli Institute for Cosmological Physics,
Laboratory for Astrophysics and Space Research, 5620 South Ellis Ave.,
Chicago, IL 60637, USA}
\affiliation{$^{27}$ Department of Physics, University of Toronto,
60 Street George Street, Toronto, ON M5S 1A7, Canada}
\affiliation{$^{28}$ Department of Physics and Astronomy, Haverford College,
Haverford, PA 19041, USA}
\affiliation{$^{29}$ Department of Physics, West Chester University
of Pennsylvania, West Chester, PA 19383, USA}
\affiliation{$^{30}$ Harvard-Smithsonian Center for Astrophysics,
Harvard University, Cambridge, MA 02138, USA}
\affiliation{$^{31}$ Department of Astronomy, University of Massachusetts,
Amherst, MA 01003, USA}

\keywords{cosmic background radiation -- cosmology: observations --
          galaxies: clusters: general -- methods: data analysis}
\maketitle

\vspace*{10cm}
\clearpage

\section{Introduction}
\label{sec_introduction}

A new generation of experiments is measuring the cosmic
microwave background (CMB) at arcminute resolutions. Within
the past year alone, results from the South Pole Telescope
\citep{staniszewski/etal:2009}, ACBAR \citep{reichardt/etal:2009},
AMiBA \citep{umetsu/etal:2009}, APEX-SZ \citep{reichardt/etal:2009a},
the Cosmic Background Imager \citep{sievers/etal:prep}, the
Sunyaev-Zel'dovich Array \citep{sharp/etal:2010}, and
QUaD \citep{friedman/etal:2009} have revealed the $\sim$arcminute
structure of the CMB with higher precision than ever.  The angular
power spectrum of temperature fluctuations at these scales
($\ell \gtrsim 1000$) will further constrain models of the early
universe. Furthermore, secondary features such as the
Sunyaev-Zel'dovich (SZ) effect and gravitational lensing probe the
growth of structure.   

With its first science release, the Atacama Cosmology Telescope (ACT)
now adds to these endeavors.  A 6\,\metre, off-axis Gregorian telescope, it
was commissioned on Cerro Toco in northern Chile in 2007 October.  Its
current receiver is the Millimeter Bolometer Array Camera (MBAC), containing
three 32$\times$32 arrays of transition edge sensor (TES) bolometers
observing at central frequencies of \arone, \artwo, and \arthree, with
beam full-widths at half-maxima (FWHM) of $1\farcm37$, $1\farcm01$,
and $0\farcm91$, respectively (see Section~\ref{ssec_beam_profiles},
below). It has operated for three seasons and is currently in its
fourth season.  In 2007 one month of science observations was made
using only the \arone\ array.  The other two frequencies were added
for the 2008 season, which lasted about 3.5 months. The telescope
optical design is described
in \citet{fowler/etal:2007}. \citet{hincks/etal:2008}
and \citet{switzer/etal:2008} report on the telescope performance and
provide an overview of hardware and software systems. The MBAC design
and details of TES detector properties and readout are
in \citet{niemack:2006}, \citet{marriage/chervenak/doriese:2006}, 
\citet{battistelli/etal:2008}, \citet{niemack/etal:2008}, 
\citet{swetz/etal:2008},   \citet{thornton/etal:2008}, 
and \citet{zhao/etal:2008}\setcounter{footnote}{31}.\footnote{Reprints
of all the references in this paragraph may freely be downloaded from:
http://www.physics.princeton.edu/act/papers.html.}

ACT is located at one of the premier sites for millimeter
astronomy because of the high altitude (5200\,m) and the dry
atmosphere.  The precipitable water vapor (PWV) had a median value
of 0.56\,mm during the nights of our 2008 season.  Nevertheless, atmospheric
emission remains the largest signal external to the receiver in our
raw data, a reality for any ground-based millimeter-wave telescope.  The
atmospheric power dominates only at low temporal frequencies and this
is the main reason we observe while scanning our telescope in
azimuth. Though much of the atmospheric power is below the frequency
of our $0.0978\,\hertz$ scans, on typical nights the atmosphere dominates
the detector noise up to about $2\,\hertz$. 

In this paper we present a map-making method designed to model and
remove the atmospheric signal in a manner which is unbiased with
respect to the celestial signal.  The method---which is independent of
the map-making pipeline used for most other ACT analysis---currently
produces its best results on small scales ($\lesssim 1\degree$), so it
is well-suited to making maps of objects with small angular sizes.
One of the most useful applications has been the study of our
instrumental beam with high signal-to-noise maps of planets.  The beam
profile affects all aspects of data analysis, including calibration,
and we provide the beam characteristics in this paper.  Additionally,
we present new SZ measurements of five known clusters.

We proceed as follows:  In Section~\ref{sec_cottingham} we introduce the
map-making method, showing both the theory and some qualitative
properties; Section~\ref{sec_beams} describes how we analyzed our beams,
and presents the key measured parameters along with beam maps
and radial profiles; window functions are derived in
Section~\ref{sec_windows}; Section~\ref{sec_clusters} shows a selection of
clusters imaged with the mapper; and we conclude in
Section~\ref{sec_conclusions}.

\section{The Cottingham Mapping Method}
\label{sec_cottingham}

In this section, we present a technique for removing the atmospheric power
first described by \citet{cottingham:1987} and used by
\citet{meyer/cheng/page:1991}, \citet{boughn/etal:1992}, and
\citet{ganga/etal:1993}.  The temporal variations in atmospheric signals are
modeled using B-splines, a class of functions ideal for interpolation,
discussed more below.  The technique computes maximum-likelihood
estimates of both the celestial and the atmospheric signals, using all
available detectors in a single frequency band.  We refer to it
hereafter as the Cottingham method.

In the following subsections, we give a mathematical description of
the Cottingham method (Section~\ref{ssec_cott_algorithm}), followed by
a discussion of its benefits and a comparison to the ``destriping''
method developed for {\em Planck}, which has close similarities
(Section~\ref{ssec_cott_comparison}).  Our approach for including the
effects of spatial variability across the detector arrays is
in Section~\ref{ssec_spatial_structure}.  We discuss the use of
B-splines in Section~\ref{ssec_bsplines}, and finish by outlining our
implementation of the method (Section~\ref{ssec_cott_implementation})
and map-making steps (Section~\ref{ssec_map_making}).

\subsection{The Algorithm}
\label{ssec_cott_algorithm}

The measured timestream $\bvec{d}$ is modeled as a celestial signal
plus an atmospheric component: 

\begin{equation}
  \bvec{d} = \bm{P}\bvec{m} + \bm{B}\bs{\alpha} + \bvec{n},
  \label{eq_signal_model}
\end{equation}

\noindent where the pointing matrix $\bm{P}$ projects the celestial
map $\bvec{m}$ into the timestream, $\bm{B}$ is a matrix of basis
functions with amplitudes $\bs{\alpha}$ which model the temporal variation of
atmospheric power, and $\bvec{n}$ is the noise.  The timestream of
measurements $\bvec{d}$ may be a concatenation of multiple detectors if
they have been properly treated for relative gain differences.
Throughout this paper, this is the case:  all working detectors from
one frequency band are processed simultaneously.

We seek $\wt{\bs{\alpha}}$ and $\wt{\bvec{m}}$, estimates of the
atmospheric amplitudes and the celestial map, respectively.
Equation~(\ref{eq_signal_model}) prescribes that we subtract the atmospheric
term to obtain the map estimate:  $\bvec{d}' = \bvec{d}
- \bm{B}\wt{\bs{\alpha}}$.  The maximum-likelihood estimator is then
given by the standard map-making
equation \citep[e.g.,][]{tegmark:1997a}:

\begin{equation}
  \wt{\bvec{m}} = \left(\bm{P}^T\bm{N}^{-1}\bm{P}\right)^{-1}
                \bm{P}^T\bm{N}^{-1}\bvec{d}'              
              = \bs{\Pi}\left(\bvec{d} - \bm{B}\wt{\bs{\alpha}}\right),
  \label{eq_map_making_eq}
\end{equation}

\noindent where we call 

\begin{equation}
  \bs{\Pi} \equiv \left(\bm{P}^T\bm{N}^{-1}\bm{P}\right)^{-1}
                  \bm{P}^T\bm{N}^{-1}
  \label{eq_def_proj}
\end{equation} 

\noindent the {\em projection matrix} and 
$\bm{N} \equiv \langle \bvec{n}\bvec{n}^T\rangle$ is the noise covariance.  The
projection matrix $\bs{\Pi}$ is designed in such a way that the map
estimate is not biased, in the sense that the error,
$\wt{\bvec{m}}-\bvec{m}$, does not depend on $\bvec{m}$.

Given a set of basis functions $\bm{B}$, the Cottingham method
minimizes the variance of the map pixel residuals with respect to the
amplitudes $\wt{\bs{\alpha}}$.  The residuals are the differences
between the celestial signals measured in the timestream and the map estimate:

\begin{align}
  \Delta\bvec{d} &= \bvec{d}' - \bm{P}\wt{\bvec{m}}
                = \bvec{d} - \bm{B}\wt{\bs{\alpha}} - 
                  \bm{P}\bs{\Pi}\left(\bvec{d}-\bm{B}\wt{\bs{\alpha}}\right)
                  \nonumber\\
               &= \left(\imat - \bm{P}\bs{\Pi}\right)
                  \left(\bvec{d} - \bm{B}\wt{\bs{\alpha}}\right),
  \label{eq_pixel_residuals}
\end{align}

\noindent where $\imat$ is the identity matrix.  We differentiate
                $\chi^2$: 

\begin{align}
  \frac{\partial\chi^2}{\partial\wt{\bs{\alpha}}}
    &= \frac{\partial}{\partial\wt{\bs{\alpha}}}
       \left(\Delta\bvec{d}^T \bm{N}^{-1} \Delta\bvec{d}\right)
       \nonumber\\
    &= -2\bm{B}^T \left(\imat - \bm{P}\bs{\Pi}\right)^T \bm{N}^{-1}
       \left(\imat - \bm{P}\bs{\Pi}\right)
       \left(\bvec{d} - \bm{B}\wt{\bs{\alpha}}\right) \nonumber\\
    &= -2\bm{B}^T \bm{N}^{-1} \left(\imat - \bm{P}\bs{\Pi}\right)
       \left(\bvec{d} - \bm{B}\wt{\bs{\alpha}}\right).
  \label{eq_cott_minimise}
\end{align}

\noindent The last equality can be obtained by expanding $\bs{\Pi}$
to its constituent elements (cf. Equation~(\ref{eq_def_proj})) and
simplifying.  If we define the following:

\begin{equation}
  \bs{\Xi} \equiv \bm{B}^T \bm{N}^{-1} \left(\imat - 
                  \bm{P}\bs{\Pi}\right), \;\;\;\;
  \bs{\Theta} \equiv \bs{\Xi}\bm{B}, \;\;\;\;
  \bs{\phi} \equiv \bs{\Xi}\bvec{d},
  \label{eq_def_xi_theta_phi}
\end{equation}

\noindent then when we set the derivative in
Equation~(\ref{eq_cott_minimise}) to zero, we have the simple expression:

\begin{equation}
  \bs{\Theta}\wt{\bs{\alpha}} = \bs{\phi}.
  \label{eq_cott_linear_eq}
\end{equation}

\noindent This is a linear equation which is straightforward to solve
for the atmospheric basis function amplitude estimates
$\wt{\bs{\alpha}}$.  These can then be used in
Equation~(\ref{eq_map_making_eq}) to estimate the map. In fact, both
$\wt{\bs{\alpha}}$ and $\wt{\bvec{m}}$ are the maximum-likelihood
estimators of the atmosphere and celestial map, respectively, for a
given set of basis functions $\bm{B}$.  We show this explicitly in
Appendix~\ref{appendix_mle}.

There is an arbitrary overall offset to the computed
$\bm{B}\wt{\bs{\alpha}}$ which must be estimated to remove the
background from maps.  We return to this point in
Section~\ref{ssec_map_making}.

\subsection{Discussion}
\label{ssec_cott_comparison}

The chief strength of the Cottingham method is that it estimates the
atmospheric power in a way that is unbiased with respect to the map
estimate itself.  This important but subtle point is encapsulated in
the term $(\imat - \bm{P}\bs{\Pi})$ in
Equation~(\ref{eq_cott_minimise}), whose effect is to project out the
map estimate from the data. Therefore, the solution to
Equation~(\ref{eq_cott_linear_eq}) is not sensitive to the estimated
celestial temperature, but only to a time-varying term which is
represented by the atmospheric estimate $\bm{B}\wt{\bs{\alpha}}$.
This should be contrasted with high-pass filtering or fitting a
slowly-varying function to the timestream to remove low-frequency
power.  Such approaches require masking of high signal-to-noise
celestial objects (such as planets or clusters) and/or multiple
iterations to prevent corruption of the maps. Simulations are required
to understand the effects of these time-domain filters on the final
maps. 

The Cottingham method has close similarities
to the ``destriping'' technique developed in particular for
{\em Planck}
analysis \citep{delabrouille:1998,burigana/etal:1999,maino/etal:2002}.
In fact, the linear algebra presented in
Section~\ref{ssec_cott_algorithm} is identical to some versions of
destriping \citep[e.g.,][]{keihanen/etal:2004}.  The
destriping techniques are intended primarily to remove $1/f$
instrumental noise---thus, for
example, \citet{keihanen/kurki-suonio/poutanen:2005} impose a
prior on the estimate $\bm{B}\wt{\bs{\alpha}}$ based on detector
noise.  \citet{sutton/etal:2009} also consider the
effects of imposing a prior on the atmospheric noise.  On the other
hand, we use the Cottingham method to remove atmospheric power with a
flat prior. A distinct feature of our method is that we process
multiple detectors simultaneously since the atmospheric signal is
common across detectors (see,
however, Section~\ref{ssec_spatial_structure}). Further, our approach 
differs in that it uses B-splines as the basis for modeling the
atmosphere (Section~\ref{ssec_bsplines}).

\subsection{Spatial Structure in the Atmosphere}
\label{ssec_spatial_structure}

The Cottingham method as presented thus far assumes that the atmospheric
signal $\bm{B}\bs{\alpha}$ is common among all the detectors.  In
fact, we know that there is also spatial structure in the atmosphere,
meaning that in principle, each detector might see a different
atmospheric signal.  In practice, the finite telescope beam sets a
lower limit on the spatial scale.  We find that the atmospheric signal
is coherent across a quarter to a third of the array, or about
5--7\,$\arcminute$.  For reference, our \arone\
channel, which has a $1\farcm37$ FWHM in
the far-field (Section~\ref{ssec_beam_profiles}), is sensitive to an angular
size of approximately $10\,\arcminute$ at a $1\,\kilo\meter$ distance,
roughly the distance to a typical turbulence layer in the atmosphere
when pointed at $50\degree$ in
altitude \citep{perez-beaupuits/rivera/nyman:2005}.  

To account for this, we divide the $32\times32$ detector array into
nine square {\em sub-arrays} of roughly equal size and fit for nine
separate temporal atmospheric signals $\bm{B}_s\bs{\alpha}_s$, with
the subscript $s$ denoting the sub-array.  These can all be done
simultaneously if we adapt Equation~(\ref{eq_signal_model}):

\begin{align}
  \bvec{d} &= \bm{P}\bvec{m} + \bm{S}
            \left(\begin{array}{cccc}
              \bm{B}_1 & \bm{0} & \hdots & \bm{0} \\
              \bm{0} & \bm{B}_2 & \hdots & \bm{0} \\
              \vdots & \vdots & \ddots & \vdots \\
              \bm{0} & \bm{0} & \hdots & \bm{B}_9 \\
            \end{array}\right)\left(\begin{array}{c} \bs{\alpha}_1 \\
                       \bs{\alpha}_2 \\ \vdots \\
                       \bs{\alpha}_9\end{array}\right)  + \bm{N}
            \nonumber\\
         &= \bm{P}\bvec{m} + \bm{S}\bm{B}'\bs{\alpha}' + \bvec{n},
  \label{eq_signal_model_brady}                              
\end{align}

\noindent where $\bm{S}$ is a book-keeping matrix that remembers
from which sub-array each measurement in $\bvec{d}$ came.  The
Cottingham method proceeds exactly as before, except that we change
$\bm{B}\rightarrow\bm{S}\bm{B}'$ and
$\bs{\alpha}\rightarrow\bs{\alpha}'$.

\subsection{The B-spline as a Model of Atmospheric Signal}
\label{ssec_bsplines}

We follow \citet{cottingham:1987} in choosing cubic basis B-splines for the
basis functions $\bm{B}$. B-splines are widely used in the field
of geometrical modeling, and numerous textbooks cover them
\citep[e.g.,][]{bojanov/hakopian/sahakian:SFAMI,deboor:APGTS:rev,schumaker:SFBT:3e};
here we summarize basic properties. Basis B-splines are a basis
of functions whose linear combination is called a B-spline.
The basis B-splines are fully determined by a knot
spacing $\tau_k$ and a polynomial order $p$; a B-spline is
flexible on scales larger than $\tau_k$, while on smaller scales
it is relatively rigid. The basis B-splines $b_{j,p}(t)$ of order $p$
are readily evaluated using the Cox--de Boor recursion on the
polynomial order.  For $m$ knots $\{t_j\}$ with $j=0$ to $m-1$:

\begin{eqnarray}
  b_{j,0}(t) &=&
  \begin{cases}
    $1$&\text{if $t_j \leq t < t_{j+1}$}\\
    $0$&\text{otherwise}
  \end{cases},
  \nonumber\\
  b_{j,p}(t) &=& \frac{t-t_j}{t_{j+p} - t_j}b_{j,p-1}(t)
  + \frac{t_{j+p+1} -t}{t_{j+p+1} - t_{j+1}}b_{j+1,p-1}(t),
\end{eqnarray}

\noindent with $j$ values restricted so that $j+p+1<m-1$.
For $m$ knot times, $m+p-1$ basis B-splines cover the interval between
the first and the last knot time. The individual basis B-splines
$b_{j,p}(t)$ are compact functions, such that the B-spline receives
support from no more than $p$ of its bases at any point. For modeling
the atmospheric signal, we always choose knots uniformly spaced in
time and use $p=3$ (cubic).  

Due to their flexibility on large scales, B-splines are ideal for
modeling the slowly varying atmospheric signal.  The frequency $f_k$ below
which power will be removed is determined by the knot spacing
$\tau_k$.  Empirically, we find:

\begin{equation}
  f_k \approx 1 / 2\tau_k.
  \label{eq_def_f_k}
\end{equation}

\begin{figure}[tb]
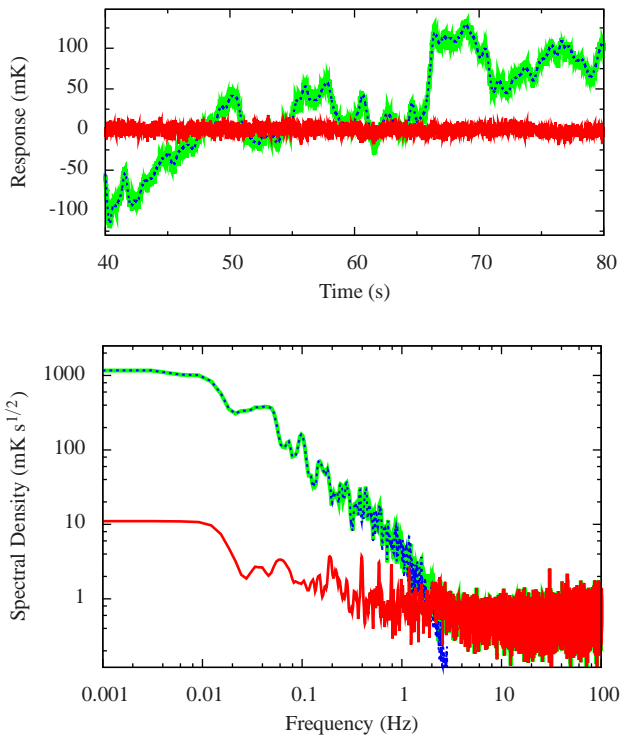

  \centering
\input{1a} \\
\input{1b}
  \caption{Example of the Cottingham method.  The fit is done
           using 300\,s of data from 605 \arone\ detectors.
           The PWV was 0.8\,mm, about
           0.25\,mm higher than the median in 2008.  The knot spacing
           is 0.25\,s and the order is cubic.  In both plots, the
           original signal is plotted with a solid, light line, the B-spline
           atmosphere model with a dashed line and the signal minus
           the model with a solid, dark line.  Each plot has been
           smoothed with a five sample boxcar filter for readability.
           The temperature units are with respect to a Rayleigh-Jeans
           spectrum. 
           Top: a portion of one of the detectors' timestreams.
           Bottom: the spectral densities for the same single 
           detector.  A Welch window was applied before computing
           the Fourier transform.}  
  \label{fig_cott_filter}
\end{figure}

Figure~\ref{fig_cott_filter} shows an example of atmospheric
estimation using the Cottingham method.  The B-spline knot-spacing is
$\tau_k = 0.25\,\second$, chosen for this example because it has $f_k
= 2\,\hertz$, the approximate frequency at which the atmospheric power
meets the detector noise level.  Longer knot spacings produce
qualitatively similar results, except that they cut off at lower
frequencies, as per Equation~(\ref{eq_def_f_k}).

The Cottingham method is effective at suppressing atmospheric
contamination, but some covariance between atmospheric and celestial
map estimates remains.  This is typically at harmonics of the scan
frequency ($\approx 0.1\,\hertz$), as exemplified in the bottom panel
of Figure~\ref{fig_cott_filter}.  For this work we have used the white
noise approximation for the detectors ($\bm{N} = 1$), which results in
maps of bright sources which are clean down to the $-40\,\deci\bel$
level in most cases (Section~\ref{sec_beams}).  The small residual
atmospheric-celestial covariance is manifested as striping along lines
of constant altitude since with our $7\degree$ peak-to-peak,
$1\fdeg5\,\second^{-1}$ azimuthal scans (or $4\fdeg47$ at
$0\fdeg958\,\second^{-1}$ when projected on the sky at our
observing altitude of $50\fdeg3$), the knot spacing ($\tau_k =
1.0\,\second$ for beam maps 
(Section~\ref{sec_beams}) and $0.5\,\second$ for cluster maps
(Section~\ref{sec_clusters})) corresponds to an angular scale smaller than
the scan width.  When noted, we fit straight lines to rows of pixels
in the map, after masking out any bright source, and subtract them.
We call this process, which takes place in the map domain, ``stripe
removal''. In both our beam analysis and our cluster studies, we have
done tests which show that the bias introduced by this process is not
significant---see Section~\ref{ssec_beam_profiles}
and Section~\ref{ssec_cluster_analysis}. Nevertheless, future extensions of
the Cottingham method would benefit from the full treatment of the
noise covariance.

\subsection{Implementation}
\label{ssec_cott_implementation}

Before making maps with the Cottingham method, some preprocessing must
be done. The data from all detectors, which are sampled at
$400\,\hertz$, are divided into fifteen-minute time-ordered data (TOD)
files and the preprocessing is performed on each individual
TOD---a future paper will describe the steps which we only summarize
here.   The data acquisition electronics' digital anti-aliasing filter
as well as measured detector time constants are deconvolved from the
raw data.  Low frequency signal due to cryogenic temperature drifts
is measured with dark detectors (i.e., detectors uncoupled to sky
signal) and removed from signal detectors; a sine wave with period
$10.23\,\second$ is also fit to each timestream and removed to reduce
scan-synchronous contamination.  Calibration to units of power uses
nightly load curves obtained by sweeping through detector bias
voltages and measuring the response. Relative gain imprecisions are
removed by using the large atmospheric signal itself to flat field the
detectors \citep[e.g.,][]{kuo/etal:2004}; this is done independently
in each of the nine sub-arrays (see Section~\ref{ssec_spatial_structure}).
Finally, calibration to temperature units uses measurements of Uranus,
for which we use a brightness temperature of $112\,\kelvin$ with 6\%
uncertainty \citep{griffin/orton:1993,marten/etal:2005,kramer/moreno/greve:2008}.
(The beam maps require no calibration to temperature---in fact, the
temperature calibration is obtained from them.)  The timestreams
require no further preprocessing.

To improve the speed of the Cottingham algorithm, we exploit the fact
that the map pixelization used for calculating the atmospheric signal
(Equations~(\ref{eq_cott_minimise})--(\ref{eq_cott_linear_eq})) need not
be the same as the map-making pixelization.  In general we only use a
selection of the possible pixels on the map; additionally, we
down-sample the number of hits in each pixel.  We call the former
``pixel down-sampling'' and denote the fraction of retained pixels
$n_p$; the latter we term ``hit down-sampling'' and denote the
fraction of retained hits $n_h$.  Consequently, the fraction of total
available data used is $n_p \times n_h$.  Each of these down-samplings
is done in an even manner such that all working detectors are used,
and also such that no large gaps exist in the remaining timestream.

We have specified four parameters for the Cottingham method:   the
knot-spacing $\tau_k$, the pixel size $\xi$, the pixel down-sampling
fraction, $n_p$, and the hit down-sampling fraction, $n_h$.  Of these,
we always choose $\xi = 18\arcsecond$ (about $1/3$ the \arthree\ beam size).

\begin{figure}[tb]
  \centering
\input{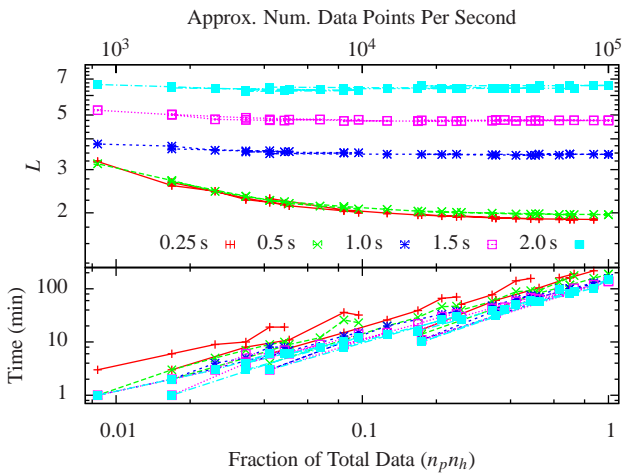}
  \caption{Top panel: the ratio $L$ of low-frequency power to
           white noise after removing the atmospheric signal, as
           defined by Equation~(\ref{eq_def_l}).  All data points are from
           the same TOD using all working detectors; only the
           parameters $\tau_k$, $n_p$ and $n_h$ (see the text) were varied
           to obtain each point.  The $x$-axis is a product of the
           resamplings $n_p$ and $n_h$, and the colors are different
           knot spacings $\tau_k$ as indicated in the plot key.
           Bottom panel: the computation time required for the
           data shown in the top panel.  (The segmented lines are an
           artifact of how the data were recorded.)  As the fraction
           $n_p n_h$ of total data used increases, the efficacy $L$ of
           power removal flattens out and adding more data does not
           significantly improve the fit but only takes more
           computation time.  In this example, there were
           $\approx103000$ possible map pixels with an average of 287
           hits pixel$^{-1}$.}
  \label{fig_drift_stat}
\end{figure}

To evaluate the effect of varying the other three variables, we define
a figure of merit which compares the average detector spectral density
below $1\,\hertz$ to the white noise level, calculated in the range
5--25$\,\hertz$:

\begin{equation}
  L \equiv \frac{1}{N_d} \left. \sum_{i}^{N_d} \left[ 
           \frac{\int_{0\,\hertz}^{1\,\hertz} G_i(f)df}
           {\int_{0\,\hertz}^{1\,\hertz}df} \middle/
           \frac{\int_{5\,\hertz}^{25\,\hertz} G_i(f)df}
           {\int_{5\,\hertz}^{25\,\hertz}df} \right] \right. ,
  \label{eq_def_l}
\end{equation}

\noindent where the sum runs over the $N_d$ detectors used for the Cottingham
calculation and $G_i$ is the spectral density of the $i$th
detector {\em after} removing the estimated atmosphere signal.
Figure~\ref{fig_drift_stat} shows a plot of measured values of $L$ for a
selection of knot-spacings and pixel down-samplings.  In our tests, we
independently varied $n_p$ and $n_h$, and found that the product 
$n_p \times n_h$, used in Figure~\ref{fig_drift_stat}, captures the
important trend in 
the ranges of interest.  As expected, shorter knot spacings remove
more power:  note however that only the $\tau_k = 0.5\,\second$ and
$\tau_k = 0.25\,\second$ are capable of removing power up to the
$1\,\hertz$ for which $L$ is defined (cf. Equation~(\ref{eq_def_f_k})).
Because the curves flatten out as $n_pn_h$ increases, at a certain
point adding more data does not substantially improve the fit.  This
supports our conclusion that we only need to use a fraction of the
data to estimate the atmosphere.

The timing data in the lower panel of Figure~\ref{fig_drift_stat} were
measured on a 64-bit Intel Xeon\textsuperscript{\textregistered}
2.5\,GHz processor.  Computation time is dominated by the calculation
of the variables in Equation~(\ref{eq_def_xi_theta_phi}).  In general,
these go linearly with the number of data points $\bvec{d}$ and
quadratically with the number of basis functions $\bm{B}$; in the case
of B-splines, the compact support of the bases can be exploited so
that the quadratic rate is subdominant to the linear, as the plot
shows. 

The number of basis B-splines is small enough that it is actually
feasible to solve Equation~(\ref{eq_cott_linear_eq}) exactly.  For
most cases, however, the conjugate gradient
method \citep[e.g.,][p. 83ff]{press/teukolsky/vetterling:NRC:2e} is
much faster and yields indistinguishable results.  Therefore, we use
the latter in our implementation. 

\subsection{Map-making}
\label{ssec_map_making}

Once the atmospheric model $\bm{B}\bs{\alpha}$ has been calculated
for a TOD with the Cottingham method, we create its celestial map.
Maps are made in $\left(\Delta a,\;\Delta A \cos(a) \right)$ coordinates,
where $\Delta a$ and $\Delta A$ are the distances from the altitude
and azimuth of the map center.\footnote{This is a very good
approximation to the Gnomonic projection for the small map sizes we
use.}  We use a pixel size of $10\farcs6$ per side, about 20\%
the size of the \arthree\ beam; note that this is different from the
pixel size $\xi$ used for calculating the atmosphere
(Section~\ref{ssec_cott_implementation}).  In this paper, the region of sky
for which the atmospheric model is calculated is much wider in the
azimuth than in altitude, since the telescope scans---$4\fdeg47$
along the $\Delta A \cos(a)$ axis---are much wider than the distance
the sky rotates along the $\Delta a$ axis as the object of interest
passes through the field of view.  However, all of the maps we
present are cropped to disks with 1:1 aspect ratios, centered on the
objects being mapped.  

We make the white noise approximation for each detector and weight it
by the inverse of its variance in the map estimate.  The detector variances
are obtained iteratively:  we make a map with equal detector weights
and measure the variances of individual detector maps against the
total map, remake the map with the new variances and repeat until the
total map variance converges.  The atmosphere estimates returned by the
Cottingham method have arbitrary offsets, which can be different for
the nine sub-arrays we use (see Section~\ref{ssec_spatial_structure}).
Thus, in the same iterative process, we also fit for the sub-array
offsets and remove them when coadding detectors.

Coaddition of TOD maps is done after all of the steps described above.
The inverse variance of each map (calculated after masking bright
point sources or clusters) is used as its median weight, and the
relative weights of its pixels are given by the number of hits per
pixel.

Finally, we mention that the software used for the results in this
paper has a completely independent pipeline from our main map-making
software which solves for the full survey area coverage.  It has been
especially useful for studying and optimizing the signal extraction in
small, targeted regions, and has provided important double-checks for
our other pipeline.

\section{Beam Maps and Properties}
\label{sec_beams}

Understanding the telescope beams, or point-spread functions, is
of primary importance for the interpretation of our maps since they
determine the relative response of the instrument to different
scales on the sky and are central to calibration.  For ACT's
measurement of angular power spectra, the Legendre transform of our
measured beam profile, called a window function, determines the
response of the instrument as a function of angular scale.

Planets are excellent sources for measuring the telescope's beam
because they are nearly point sources and are brighter than almost
any other celestial object.  The best candidates for ACT are Saturn
and Mars; of the rest, Jupiter is too bright and saturates the
detectors, Venus is available too near to sunrise or sunset when the
telescope is thermally settling, and the others are too dim for
exploring the far sidelobes of the beam. (However, Uranus is useful as
a calibrator---see Section~\ref{ssec_cott_implementation}.) The beam
maps presented in this section are from observations of Saturn, which
was available from early November through December of 2008.  The
possible concern of detector non-linearities when observing Saturn is
obviated by the fact that atmospheric fluctuations, which are of the
same order of magnitude in brightness as Saturn, are found to produce
linear detector responses.

\subsection{Data Reduction}
\label{ssec_beams_reduction}

Maps were made for each night-time TOD of Saturn, using the Cottingham
method with $\tau_k = 1\,\second$, $n_p = 0.32$ and $n_h
= 0.36$.  We had more TODs than were needed to make low-noise beam
maps, and we excluded about 1/3 of the maps which had higher
residual background contamination, manifested by beam profiles that
significantly diverged from those of the cleaner maps.  The map sizes
and number of TODs per frequency band are shown in
Table~\ref{tab_beam_summary}.

\begin{figure*}[t]
  \centering
\input{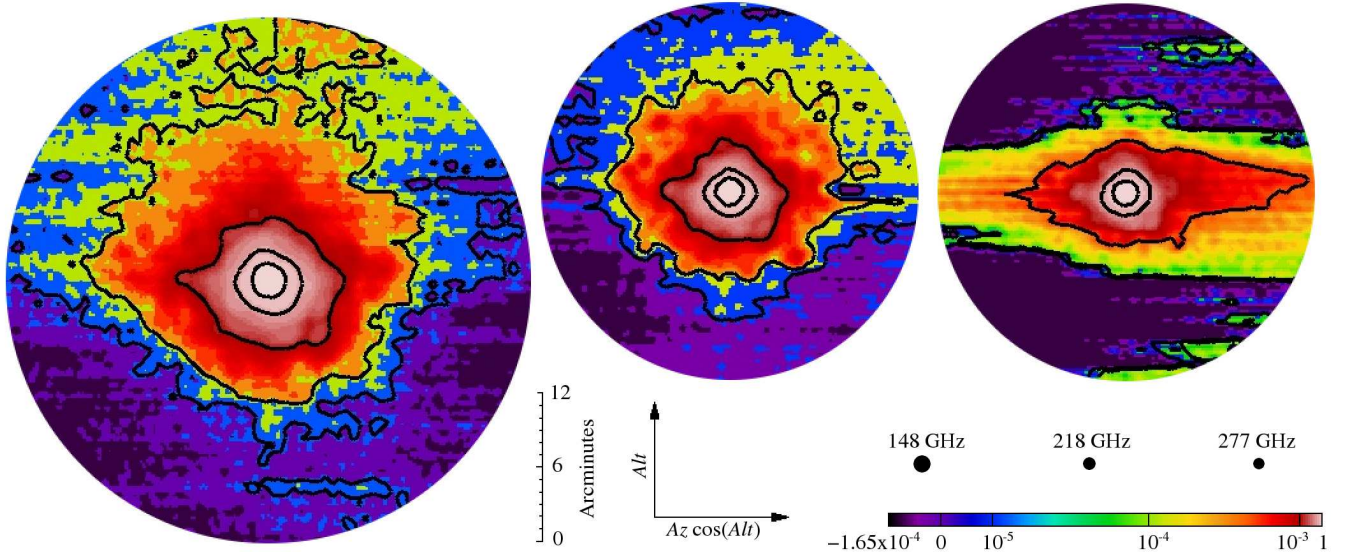}
  \caption{Beam maps for the three frequency bands:  from left to
           right, \arone, \artwo\ and \arthree.  The maps are from
           coadded observations on 11--15 nights (see
           Table~\ref{tab_beam_summary}) and have radii of
           $21\arcminute$ (\arone) and $15\arcminute$ (\artwo\
           and \arthree).  Maps are normalized to unity and contours
           are in decrements of $-10$\,dB.  A histogram equalized
           color scale is used to highlight the fact that we
           have made $< -40\,\deci\bel$ beam measurements of our \arone\
           and \artwo\ bands.  (Negative values are due to noise
           about the mean background level.)  Even in the
           inferior \arthree\ map, striping is still below
           $-20\,\deci\bel$.  The disks above the color bar show the
           sizes of the beam FWHM for each band (see
           Table~\ref{tab_beam_summary}).  A Gaussian smoothing kernel
           with $\sigma = 0\farcm54$ has been applied to highlight
           large-scale structure; smoothing is not otherwise performed
           in the analysis.  No stripe removal has been done on these
           maps.} 
  \label{fig_wing_maps}
\end{figure*}

In the analysis of Section~\ref{ssec_beam_profiles}, below, we sometimes
compare unprocessed maps to stripe-removed maps.  Stripe-removed map
have been treated as outlined in Section~\ref{ssec_bsplines}, while
unprocessed maps have not been altered after map-making except for
the subtraction of a background level. The planet is masked out before
calculating the mean map value which estimates this level.  The mask
sizes for the three arrays, whether used for stripe removal or
background estimation, are listed in Table~\ref{tab_beam_summary}.

In each frequency band, the selected TOD beam maps were coadded.
Weights were determined from the rms of the mean background level,
calculated outside the mask radius.   Relative pointing of individual
detectors was measured to sub-arcsecond precision using the ensemble
of Saturn observations.  The overall telescope pointing was
determined from each planet observation prior to map-making and used
to center each TOD map, so recentering of the maps was unnecessary
before coaddition.
\vspace{1cm}

\subsection{Beam Measurements}
\label{ssec_beam_profiles}

Figure~\ref{fig_wing_maps} shows coadded beam maps for the three arrays
using a color scale which highlights the features in the sidelobes.
The \arone\ and \artwo\ maps have striking similarities,
most notably along the altitude (or vertical) direction where both
exhibit more power near the top of the map.  This is due to the
off-axis design of the telescope \citep[see][]{fowler/etal:2007}: since the
\arone\ and \artwo\ arrays sit at about the same vertical
offset from the center of the focal plane, their resemblance along
this axis is expected.  Note that we recover structure in these map at a
$< -40\,\deci\bel$ level.  The \arthree\ map is clearly inferior,
showing residual striping in the scan direction, although this occurs
below $-20\,\deci\bel$.  We believe that this is from a combination of the
brighter atmosphere at \arthree, as well as detector noise correlation
induced by large optical loads (such as Saturn), which we are still
investigating.  Nonetheless we are still able to measure the \arthree\
solid angle to about 6\% (see below), and work is underway to improve
it.

For the \arone\ and \artwo\ arrays, we do our beam analysis on maps
which have not had stripe removal because this process removes the
real vertical gradient from the maps (see Figure~\ref{fig_wing_maps}).
Nonetheless, the solid angles (see below) from stripe-removed
maps are within $1\sigma$ of the values from unprocessed maps.  On the 
other hand, the larger residual striping in the \arthree\ maps
necessitates the use of stripe-removed maps.

The beam center is characterized by fitting an elliptical Airy
pattern---the function describing the beam of an optical system with a
perfect aperture---to the top $\sim$3\,dB of the beam map.  This
provides a measurement of the location of the beam center, its FWHM
along the major and minor axes of the ellipse, and its orientation,
which we define as the angle of the major axis from the line of zero
altitude relative to the beam center.  The uncertainties in these
parameters are determined using the bootstrap
method  \citep[pp. 691ff]{press/teukolsky/vetterling:NRC:2e} and give
errors consistent with the standard deviation of values measured from
the individual TOD maps.  The FWHM and angles are listed in
Table~\ref{tab_beam_summary}.  They are included for reference but
are not used in any analysis. 

\begin{deluxetable}{lccc}
  \tablecaption{Summary of Beam Parameters}
  \tablewidth{0pt}
  \tablehead{
    & \colhead{\arone} & \colhead{\artwo} & \colhead{\arthree}
  }
  \startdata
    \multicolumn{4}{l}{
      Map properties (Section~\ref{ssec_beams_reduction})
    }\\
    \;\;\;No. of TODs & 16 & 15 & 11 \\
    \;\;\;Stripe removal? & no & no & yes \\
    \;\;\;Map radius ($\arcminute$) & 21 & 15 & 15 \\
    \;\;\;Mask radius ($\arcminute$)\tablenotemark{a} & 18 & 9,11,13 & 
         6,8,10,12 \\
    &&&\\
    \multicolumn{4}{l}{
      Beam centers (Section~\ref{ssec_beam_profiles})
    }\\
    \;\;\;Major FWHM ($\arcminute$) & $1.406 \pm 0.003$ & 
                              $1.006 \pm 0.01$ & $0.94 \pm 0.02$ \\
    \;\;\;Minor FWHM ($\arcminute$) & $1.344 \pm 0.002$ & 
                              $1.001 \pm 0.003$ & $0.88 \pm 0.02$ \\
    \;\;\;Axis angle ($\degree$) & $62 \pm 2$ & $137 \pm 9$ &
                                         $98 \pm 13$ \\
    &&&\\
    \multicolumn{4}{l}{
      $\theta_W$ wing fits (Section~\ref{ssec_beam_profiles})
    }\\
    \;\;\;Fit start, $\theta_1$ ($\arcminute$) & 7 & 5 & 4.5 \\   
    \;\;\;Fit end, $\theta_2$ ($\arcminute$)\tablenotemark{b} & 
                                           13 & 7--11 & 6--10 \\
    \;\;\;Best-fit $\theta_W$ ($\arcminute$) & $0.526\pm 0.002$ &
                                   $0.397 \pm 0.01$ & $0.46 \pm 0.04$ \\
    &&&\\
    \multicolumn{4}{l}{
      Solid angles (Section~\ref{ssec_beam_profiles})
    }\\
    \;\;\;Solid angle (nsr) & $218.2 \pm 4$ & $118.2 \pm 3$ & $104.2 \pm 6 $ \\
    \;\;\;Percent interpolated & 2.8 & 4.3 & 7.2\\
    &&&\\
    \multicolumn{4}{l}{
      Beam fits (Section~\ref{ssec_window_bases})
    }\\
    \;\;\;$\theta_0$ ($\arcminute$) & 0.2137 & 0.1562 & 0.1367 \\
  \enddata
  \tablecomments{See the text for definitions of these parameters and how
                 they are measured.  Values for the \arone\ and
                 \artwo\ bands are obtained from the coadded,
                 unprocessed maps whereas the \arthree\ values are
                 average values from stripe-removed maps with the mask
                 sizes indicated in the table.}
  \tablenotetext{a}{The \artwo\ and \arthree\ beam
                    properties are averaged from the results at 
                    these mask radii---see the text.} 
  \tablenotetext{b}{The fit ranges for the \artwo\ and \arthree\ band
                    are varied along with the mask radii so that
                    $\theta_2$ is never larger than the mask---see
                    text.}
  \label{tab_beam_summary}
\end{deluxetable}

We denote the beam map by $B(\theta,\phi)$, where we use coordinates
with radial distance $\theta$ from the beam center and polar angle
$\phi$.  By definition, $B(0, \phi) = 1$.  The symmetrized beam is
averaged around the polar angle: 

\begin{equation}
  b^S(\theta) \equiv \frac{\int d\phi' B(\theta, \phi')}
                     {\int d\phi'}.
  \label{eq_def_beam_profile}
\end{equation}

Another quantity of interest is the accumulated solid angle, which
measures the total normalized power within a given radius:

\begin{equation}
  \Omega(\theta) = \int\limits_0^{2\pi}d\phi'
                   \int\limits_0^{\theta}\theta' d\theta' 
                   B(\theta', \phi').
  \label{eq_def_accum_sa}
\end{equation}

\noindent The beam solid angle  is 
$\Omega_A \equiv \Omega(\theta = \pi)$.

\begin{figure*}[htb]
  \centering
\input{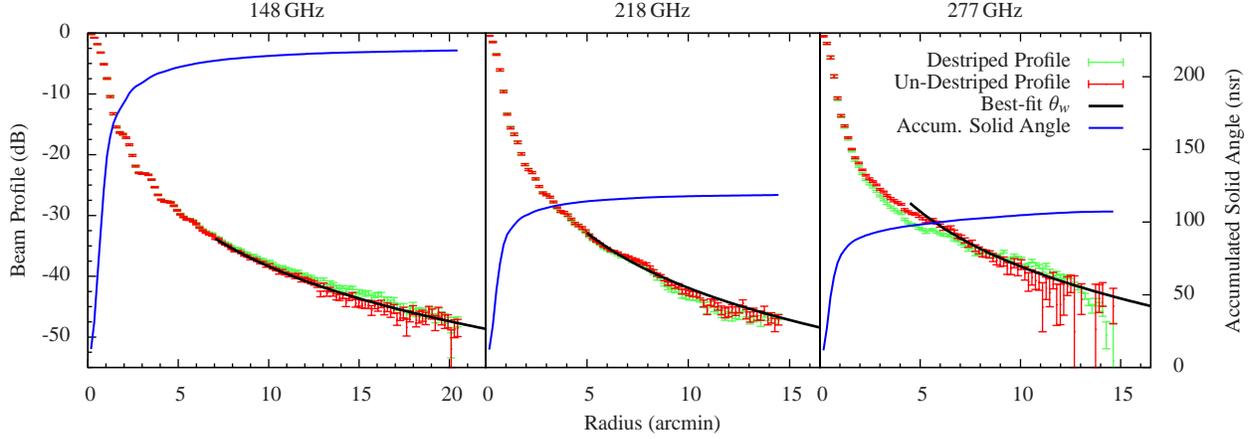}
  \caption{The beam profiles (Equation~(\ref{eq_def_beam_profile})) and
           accumulated solid angles (Equation~(\ref{eq_def_accum_sa}))
           for the three arrays, calculated from coadded maps (see
           text).  The beam profiles are shown for both unprocessed
           maps, with dark errorbars, and stripe-removed maps, with
           light errorbars. Over-plotted on each profile is the best
           fit of $\theta_W$ (Equation~(\ref{eq_def_theta_w})) to the
           unprocessed beam profiles.  The error on the profiles are
           standard errors from the azimuthal average.  The
           accumulated solid angles are from the unprocessed maps
           (without any solid angle extrapolation via
           Equation~(\ref{eq_missing_sa})) for \arone\ and \artwo, and
           from the stripe-removed map for \arthree.  Saturn is bright
           enough that the rms power from the CMB falls below all the
           points in these plots.}  
  \label{fig_profiles}
\end{figure*}

Figure~\ref{fig_profiles} shows measured beam profiles and accumulated
solid angles for the three arrays.  We measure the beam profiles down
to about $-45\,\deci\bel$.  If the beams exactly followed an Airy
pattern, these data would account for 98\% of the solid angle.
Since systematic effects could corrupt our maps at the largest radii,
we seek a way to robustly estimate the last few percent of the solid
angle on each beam.  The method is to extrapolate the data with a fit
to the asymptotic expression for the Airy pattern:

\begin{equation}
  b^S(\theta \gg \theta_F) =
  \left(\frac{\theta_W}{\theta}\right)^3,
  \label{eq_def_theta_w}
\end{equation}

\noindent where $\theta_F$ is the beam FWHM and $\theta_W$ defines the
wing scale. Equation~(\ref{eq_def_theta_w}) is good to better than 1\%
beyond about $5\theta_F$ \citep[Section~10.2b]{schroeder:AO:2e}.\footnote{In
full generality Equation~(\ref{eq_def_theta_w}) is also proportional
to $\cos(\pi D \theta/\lambda - 3\pi/4)$, where $D$ is the telescope
aperture diameter and $\lambda$ is the wavelength; we have smoothed over
cosine cycles.} Knowledge of $\theta_W$ allows us to infer the amount
of unaccounted solid angle beyond the map boundary.  A simple
integration shows that the solid angle beyond a radius $\theta_b$ is:

\begin{equation}
  \Omega_W(\theta > \theta_b) = 2\pi\frac{\theta_W^3}{\theta_b}.
  \label{eq_missing_sa}
\end{equation}

We can also use this expression to estimate the amount of true beam
power which was ``mistakenly'' included in the measurement of the
background level outside the mask radius and subtracted from the map.
In our analysis of the beam profiles and solid angles (including those
displayed in Figure~\ref{fig_profiles}), we use the fits of $\theta_W$
to calculate this missing power and add it back into the map.  A new
$\theta_W$ is then calculated from the corrected map; after two such
iterations the $\theta_W$ fit converges.

Figure~\ref{fig_profiles} includes over-plots of the wing estimates from the
best-fit values of $\theta_W$ on unprocessed maps.  We denote the
radii between which the fits were performed as $\theta_1$--$\theta_2$,
and choose $\theta_1 \approx 5\theta_F$ for each array.  For \arone, we
obtain good fits for any choice of $\theta_2$ up to $13\arcminute$,
or about the $-40\,\deci\bel$ level in the profile.  Thus, we use
$\theta_2 = 13\arcminute$, for which we fit with $\chi^2$ of 40 for 35
degrees of freedom.  The fits to the other profiles are not as
robust:  \artwo\ has a reduced-$\chi^2$ of $2.8$ for
$\theta_2 = 7\arcminute$ and \arthree\ has reduced-$\chi^2$ of $25$ for
$\theta_2 = 6\arcminute$. Larger $\theta_2$ gave poorer fits.
Consequently, for these profiles we calculate $\theta_W$ at different
mask sizes, as indicated in Table~\ref{tab_beam_summary}.  At each mask size
we varied $\theta_2$ in $2\arcminute$ increments, always keeping it lower than
the mask size.  The average value from the whole ensemble of fits gives us
$\theta_W$ and we take its standard error as the uncertainty.
Although Equation~(\ref{eq_def_theta_w}) may be too simple a model for these
profiles, contributions to the solid angle at these radii are only a
few percent of the total solid angle, which has an uncertainty dominated
by the contribution of the beam at radii less than $\theta_2$---see
below.  The values of $\theta_W$ for all three beam profiles are listed
in Table~\ref{tab_beam_summary}. 

Our $\theta_W$ fits allow us to calculate precise solid angles.  At
radii smaller than $\theta_2$, we integrate the normalized power in
the map (cf. Equation~(\ref{eq_def_accum_sa})).  Beyond $\theta_2$, we use
Equation~(\ref{eq_missing_sa}) to extrapolate the remaining solid angle.  (In
the case of the \artwo\ and \arthree\ solid angles, we use the
smallest $\theta_2$ and the largest mask size in the ranges shown in
Table~\ref{tab_beam_summary}.  Other choices from these ranges do not
significantly alter the results.)  Finally, in the approximation that
Saturn is a solid disk, it adds half of its solid angle $\Omega_S$ to
the measured instrument solid angle---this is shown in
Appendix~\ref{appendix_sa}.  Thus, the total solid angle is: 

\begin{equation}
  \Omega_A \;=\; \Omega(\theta \le \theta_2) \;+\; \Omega_W(\theta >
                 \theta_2) \;-\; \Omega_S / 2.
  \label{eq_tot_sa}
\end{equation}

\noindent During the period of our observations, Saturn subtended
solid angles from 5.2 to 6.0 nanosteradians (nsr).  We use
the mean value of 5.6\,nsr.

Determining the rest of the uncertainty in the solid angle is not
straightforward since systematic errors dominate.  For our total
error, we add the estimated uncertainties of each of the terms on the
right-hand side of Equation~(\ref{eq_tot_sa}) in quadrature.  The
uncertainty from Saturn's solid angle we take to be 1\,nsr, both
because of its varying angular size and to account for any systematic
error due to the disk approximation.\footnote{The rings of Saturn add
a layer of complication to its solid angle calculation, particularly
since they have a different temperature than the disk.  The ring
inclination was low during our observations ($< 6\degree$) and we have
estimated that their contribution is negligible within the error
budget.}  The uncertainty of $\Omega_W$ is derived from the error of
the fitted $\theta_W$.  For $\Omega(\theta < \theta_2)$, which
dominates, we estimate the error by looking at the distribution of
values from the individual TOD maps which comprise the coadded map. We
did this in two ways. First, we calculated the mean and standard
deviation of the solid angles measured in each individual map.  This
also reassures us that the coaddition step does not introduce any
systematic error through, for example, pointing misalignments or
changes in telescope focus from night to night. Second, we used the
bootstrap method to generate 1000 coadded maps with random subsets of
individual maps and used this ensemble to estimate the 68th percentile
(i.e., $1\,\sigma$) of solid angles.  These two error estimates were
consistent with each other.

The solid angles and their uncertainties are reported in
Table~\ref{tab_beam_summary}.  The formal uncertainties have
been doubled and we quote them as $1\,\sigma$, in case there are
systematic effects for which we have not accounted.  In particular,
the maps used for power spectrum estimation will come from an
independent pipeline and will treat the instrumental
response in slightly different ways---for example, by weighting
detectors differently.  We expect the beam uncertainties to decrease
as our analysis evolves.

\section{Window Functions}
\label{sec_windows}

The statistics of the CMB are frequently characterized by an angular
power spectrum $C_{\ell}$:

\begin{equation}
  \Delta T(\bh{n}) = \sum_{\ell,m}a_{\ell m}Y_{\ell m}(\bh{n});\;\;\;
  \left\langle a^*_{\ell'm'}a_{\ell m} \right\rangle =
  \delta_{\ell'\ell}\delta_{m'm}C_{\ell},
  \label{eq_ang_pow_spectrum}
\end{equation}

\noindent where $\Delta T(\bh{n})$ is the CMB temperature at position
$\bh{n}$ and $Y_{\ell m}$ is a spherical harmonic. In spherical
harmonic space, the beam is encoded in a window function
$w_{\ell}$ describing the response of the experiment to different
multipoles $\ell$, such that the total variance of a noiseless power
spectrum is:

\begin{equation}
  \mathrm{Var} = \sum_{\ell} \frac{2\ell + 1}{4\pi}C_{\ell}w_{\ell}.
\end{equation}

In the case of a symmetric beam, the window function is
the square of the Legendre transform of the beam radial profile
\citep{white/srednicki:1995,bond:1996}:

\begin{equation}
  w_{\ell} = b_{\ell}^2;\;\;\;
  b_{\ell} \equiv \frac{2\pi}{\Omega_A}
  \int b^S(\theta)P_{\ell}(\theta)d(\cos\theta).
  \label{eq_legendre_transform}
\end{equation}

\subsection{Basis Functions}
\label{ssec_window_bases}

For calculation of the window function and its covariance we
model each beam with a set of basis functions which is complete but
not necessarily orthogonal:

\begin{equation}
  b^S(\theta) = \sum_{n=0}^{n_{\rm max}} a_n b_n(\theta).
  \label{eq_beam_basis_expansion}
\end{equation}

Because the beam is truncated by a cold Lyot stop
\citep{fowler/etal:2007}, its Fourier transform is compact on a disk,
which suggests that a natural basis with which to decompose the Fourier
transform of the beam image is the set of Zernike polynomials that
form an orthonormal basis on the unit
disk \citep{born/wolf:POO:7e}. The Zernike polynomials, expressed in
polar coordinates $\rho$ and $\varphi$ on the aperture plane, are:

\begin{equation}
  V^m_n(\rho, \varphi) = R^m_n(\rho) e^{i m \varphi},
  \label{eq_def_zernikes}
\end{equation}

\noindent where $m$ and $n$ are integers such that $n \geq 0,$ $n >
|m|$ and $n-|m|$ is even. In the case of an azimuthally symmetric
beam, we need only consider the $m=0$ radial polynomials, which can be
expressed in terms of Legendre polynomials, $P_n(x)$, as follows:

\begin{equation}
  R^0_{2n}(\rho) = P_n(2\rho^2-1).
  \label{eq_symmetric_zerikes}
\end{equation}

\noindent The radial Zernike polynomials have a convenient analytic
form for their Fourier transform:

\begin{equation}
  \tilde{R}^0_{2n}(\theta) = \int\rho d\!\rho\,e^{-i \rho \theta }
  {R}^0_{2n}(\rho) = (-1)^{n} J_{2n+1}(\theta)/\theta,
  \label{eq_zernike_ft}
\end{equation}

\noindent where $J_n$ is a Bessel function of the first kind.
Motivated by this, we adopt

\begin{equation}
  b_n(\theta) = \left(\frac{\theta}{\theta_0}\right)^{-1}
  J_{2n+1}\left(\frac{\theta}{\theta_0}\right)
  \label{eq_beam_basis}
\end{equation}

\noindent as our set of basis functions to fit the radial beam
profile.\footnote{It may be asked why
the Airy pattern, which was somewhat suitable for the high-$\theta$ fit
in Section~\ref{ssec_beam_profiles}, is not used here. We find that at
low $\theta$, it is a poor fit since the optics are more complicated
than the perfect-aperture model assumed by the Airy pattern.}  Here,
we have introduced a fitting parameter, $\theta_0,$ to control the
scale of the basis functions. 

\subsection{Fitting Basis Functions to the Beam Profile}
\label{ssec_beam_fit}

Below $\theta_1$ (cf. Table~\ref{tab_beam_summary}), we fit the bases
$b_n$ of Equation~(\ref{eq_beam_basis}) to the measured beam profile, and
beyond $\theta_1$, we use the power law defined in
Equation~(\ref{eq_def_theta_w}) with the parameters $\theta_W$ listed in
Table~\ref{tab_beam_summary}. We assume vanishing covariance between
the power law and the basis functions as they are fitted to
independent sets of data points.

We employ a nonlinear, least-squares method
to solve for the coefficients $a_n$ and their covariance matrix
$C^{aa'}_{mn}$. The algorithm uses a singular value decomposition to
determine if the basis functions accurately characterize the data and
also computes a goodness-of-fit statistic
\citep[Section~15.4]{press/teukolsky/vetterling:NRC:2e}. As inputs to the
fitting procedure we are required to specify the scale parameter,
$\theta_0,$ and polynomial order, $n_{\rm max}.$ We searched the
$\{\theta_0,n_{\rm max}\}$ parameter space until a reasonable fit was
obtained that kept $n_{\mathrm{max}}$ as small as
possible. For all three bands, $n_{\mathrm{max}} = 13$ gives a reduced
$\chi^2 \approx 1$. No singular values is found for any of the
fits. The parameters $\theta_0$ we use for each frequency band are
listed in Table~\ref{tab_beam_summary}.

\subsection{Window Functions and Their Covariances}

Given the amplitudes $a_n$ of the radial beam profile fitted to the
basis functions and the covariance matrix $C^{aa'}_{mn}$ between the
amplitudes $a_m$ and $a_n$, the beam Legendre transform is:

\begin{equation}
  b_\ell = \sum_{n=0}^{n_{\rm max}} a_n b_{\ell\,n}.
\end{equation}

\noindent and the covariance matrix of the beam Legendre transforms $b_{\ell}$
and $b_{\ell'}$ is:

\begin{equation}
  \Sigma^b_{\ell \ell'} =
  \sum_{m,n=0}^{n_{\rm max}} \frac{\partial b_\ell}{\partial a_m}
  C^{aa'}_{mn} \frac{\partial b_{\ell'}}{\partial a_n}
\end{equation}

\noindent and therefore the covariance for the window function is:

\begin{equation}
  \Sigma^w_{\ell \ell'} = 4 w_\ell w_{\ell'} \Sigma^b_{\ell \ell'}.
  \label{eq_window_cov}
\end{equation}

\begin{figure}[tb]
  \centering
\input{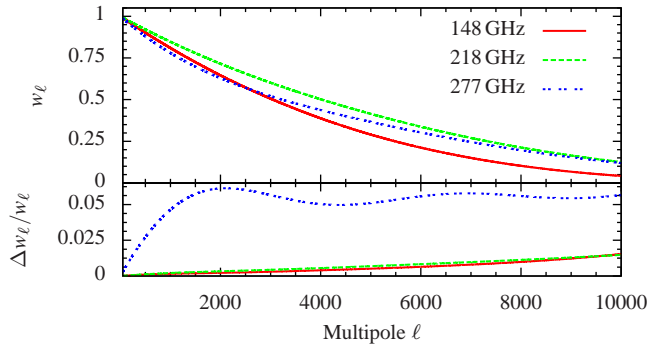}
  \caption{Normalized window functions (top) and diagonal errors
           (bottom) computed from the basis functions for each
           of the three frequency bands.   The window functions have been
           normalized to unity at $\ell=0$. In practice, the
           normalization will take place over the range of multipoles
           corresponding to the best calibration. Only statistical
           errors are shown.} 
  \label{fig_windows}
\end{figure}

In Figure~\ref{fig_windows} we show the window functions for each of
the three frequency bands with diagonal error bars taken from the
covariance matrix, $\Sigma^w_{\ell \ell'}$. We observe that the window
function for each of the frequency bands has fallen to less than
$15\%$ of its maximum value at $\ell=10000$.  The statistical diagonal
errors are at the $1.5\%$, $1.5\%$, and $6\%$ levels for the \arone, \artwo,
and \arthree\ bands respectively, as shown in Figure~\ref{fig_windows}.
They are computed following Equation~(17) of \citet{page/etal:2003}.
The off-diagonal terms in the beam covariance matrix are comparable in
magnitude to the diagonal terms. Singular value decompositions of
$\Sigma_{\ell\ell'}^b$ yield only a handful of modes with singular
values larger than $10^{-3}$ of the maximum values:  five modes
for \arone, four for \artwo, and two for \arthree. Thus, the window
function covariances can be expressed in a compact form which will be
convenient for power spectrum analyses.

For \arthree\ we estimate a $10\%$ systematic uncertainty from
destriping.  Another source of systematic error in the window
functions arises from the beams not being perfectly symmetric.  The
symmetrized beam window function generally underestimates the power in
the beam  \citep[e.g., Figure~23,][]{hinshaw/etal:2007}.  In practice,
the scans in our survey field are cross-linked
(see Section~\ref{ssec_cluster_data}).  Thus, given the spherical
transform of a cross-linked beam map $b_{\ell}$, the window function
of relevance for the power spectrum calculation is: 

\begin{equation}
  w_\ell= {1\over 2\pi} \int^{2\pi}_{0} b(\bs{\ell})\,
  b^{*}(\bs{\ell})\,d\phi_{\ell},
  \label{eq_cross_linked_win}
\end{equation}

\noindent where $\phi_{\ell}$ is the polar angle in spherical harmonic space.
We computed the fractional difference between the window function
derived from the Legendre transform of the symmetrized beam (shown in
Figure~\ref{fig_windows}) to that derived from an estimated two-dimensional,
cross-linked beam map.  For the latter, we rotated our Saturn beam map
by a typical cross-linking angle of $60\degree$, coadded it with the
original, and evaluated Equation~(\ref{eq_cross_linked_win}) on the
transform of this synthetic map.  The difference between the two was
found to be at the $1\%$ level for the \arone\ and \artwo\ arrays, and
at the $4\%$ level for the \arthree\ array.

\section{SZ Galaxy Clusters}
\label{sec_clusters}

In addition to beam maps, the Cottingham method mapper has been used for
making maps of SZ clusters.  The maps and analysis presented in this
section are the first results from ACT on SZ science.  For this first
overview, we include results from only the \arone\ band, the most
sensitive during our 2008 season.  Of the five clusters presented in
this paper, all of which were previously known, three are detected for
the first time with the SZ effect.

\subsection{Data}
\label{ssec_cluster_data}

Table~\ref{tab_clusters} lists the clusters studied in this
paper, including information on the maps and a summary of the results
of our analysis
(Section~\ref{ssec_profiles}, Section~\ref{ssec_cluster_analysis}).
Figure~\ref{fig_clusters} shows the cluster maps and companion
difference maps (see below).

\begin{figure}[t!]
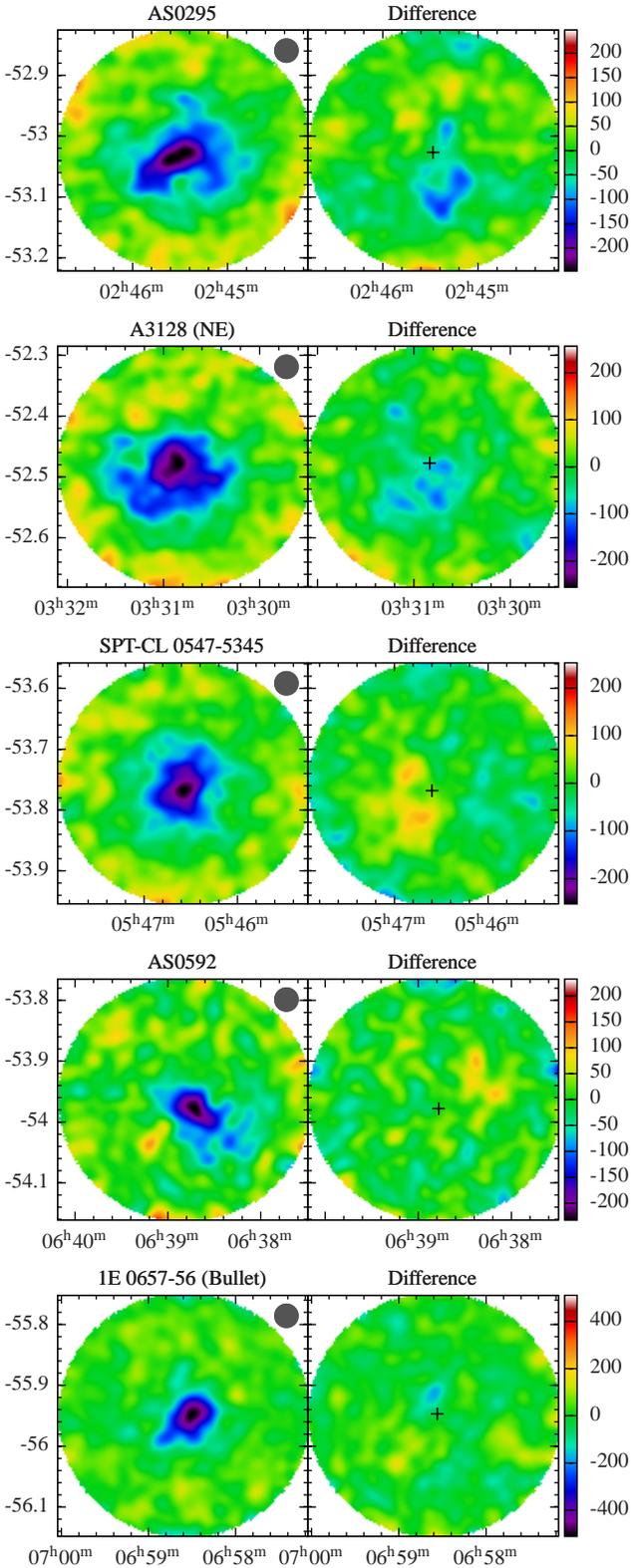

  \center
\input{6a}
\input{6b}
\input{6c}
\input{6d}
\input{6e}
  \caption{Cluster maps made using the Cottingham method at \arone,
           paired with their difference maps
           (see Section~\ref{ssec_cluster_data}).  The coordinates are
           J2000 right ascension (hours) and declination (degrees).
           The color bars are $\micro\kelvin$ (CMB); note that the scale
           is different for each cluster. The gray disk in the top
           corner of the the signal plots is $2\farcm43$ in
           diameter, the FWHM size of the beam convolved with the
           Gaussian smoothing kernel which was applied to these
           images.  In each difference plot, a cross shows the
           coordinates of the darkest spot in its corresponding signal
           map.} 
  \label{fig_clusters}
\end{figure}

Apart from planets, ACT has done no targeted observations of
specific objects, so the cluster maps come from our regular survey
data, which were taken at two different central azimuth pointings, one on
the rising sky and the other on the setting sky.  Therefore, the maps
presented here are ``cross-linked'', i.e., they consist of data taken
with two distinct angles between the azimuthal scan direction and the
hour angle axis.  The integration time is short, ranging from about
3 to 11 minutes---see Table~\ref{tab_clusters}.

The clusters were found in a full-survey \arone\ map produced
by our main map-maker.  A Wiener filter was
constructed using the polytropic model of \citet{komatsu/seljak:2001}
as an SZ template, and included detector noise, CMB power and point
source contributions in the noise model.  Clusters were then
identified from the filtered maps. We make two points about the
clusters presented in this paper: first, although our template-based
detection method has some built-in bias, the detections presented here
are significant ($\geq 5\sigma$ in the filtered survey map); and
second, we have only included a sample of our significant 
detections.

Cluster maps are made using the procedures outlined in
Section~\ref{ssec_cott_implementation} and
Section~\ref{ssec_map_making}. The knot spacing was $\tau_k =
0.5\,\second$ and the downsampling fractions were $n_p \approx 0.42$
and $n_h \approx 0.40$.  All maps are $0\fdeg4$ in diameter.
Straight-line stripe removal (Section~\ref{ssec_map_making}) has been
performed, using a $6\arcminute$ radius mask over the cluster
decrement.

We have made companion ``difference'' maps for each cluster from the same
data.  For each of the rising and setting observations, a map made
from the first half of the season's data is subtracted from the second
half.  The rising and setting difference maps are then coadded to produce
the full, cross-linked coadded difference map---the same procedure used for
the signal maps.

The map noise, listed in Table~\ref{tab_clusters}, is the rms of the
map computed outside a $6\arcminute$ mask and converted to an
effective pixel size of 1\,arcmin$^2$.  By examining the power
spectra of the maps we found that the rms values we quote are
dominated by the white noise level and do not have significant
contributions from residual low-frequency power.

\subsection{Cluster Profiles}
\label{ssec_profiles}

The cluster center positions are determined by finding the coldest 
point in the map smoothed with a $2\arcminute$ FWHM Gaussian
kernel.  
As a rough guide, we also quote the cluster depths, $\Delta
T_{\mathrm{SZ}}$, from these smoothed maps in
Table~\ref{tab_clusters}, but we stress that these values should not
be used for quantitative analysis.  All other cluster properties are
measured from unsmoothed maps.

\begin{figure*}[htb]
\input{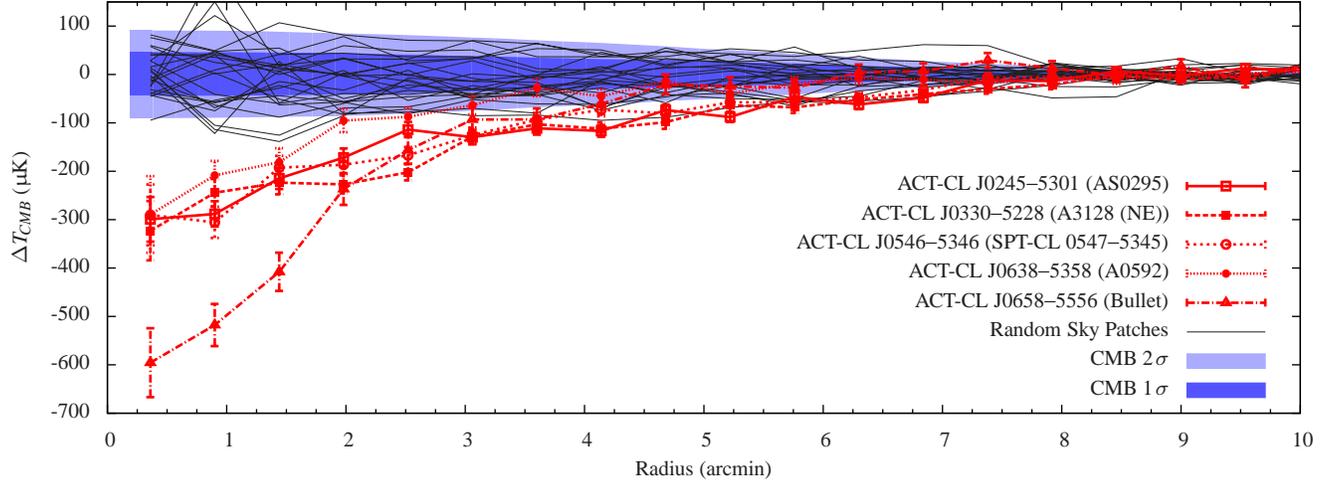}
  \caption{Radial cluster profiles, calculated by finding the mean
           temperature in the maps in $32\arcsecond$~wide annuli.  For
           comparison, the radial profiles for the 24 maps of random
           patches of the sky, not containing clusters, are included;
           additionally, shaded areas show the range containing 68\% and
           95\% of profiles from an ensemble of 1000 simulated,
           noise-free CMB patches
           (cf. Section~\ref{ssec_cluster_analysis}).  The locations
           of the profile centers are listed in Table \ref{tab_clusters}.
           The $1\farcm4$ FWHM beam has not been deconvolved from
           any of the profiles.} 
  \label{fig_sz_profiles}
\end{figure*}

Figure~\ref{fig_sz_profiles} shows the clusters' radial temperature
profiles, calculated by finding the mean map temperatures in
$32\arcsecond$~wide annuli about the cluster centers listed in
Table~\ref{tab_clusters}.  The errorbars on the profiles plotted in
Figure~\ref{fig_sz_profiles} are the standard errors of these mean
values.

To highlight the size of the SZ decrements, 24 additional profiles are
included in Figure~\ref{fig_sz_profiles}, each from a map of a patch
of the sky containing no clusters and processed in the same way as the
cluster maps.  Furthermore, to illustrate that the depth of the
cluster profiles is much greater than the fluctuations in the primary
CMB anisotropies, we created a large ensemble (1000) of noise-free,
primary-anisotropy CMB simulations using a flat $\Lambda$CDM
cosmology.  These synthetic maps were made at the same size and
resolution as our signal maps, and underwent the same stripe removal
process. Figure~\ref{fig_sz_profiles} shows shaded areas which contain
68\% and 95\% of the simulated CMB radial profiles.  The cluster
profiles fall significantly below both the simulated CMB-only profiles
and the profiles from blank CMB patches.

As a check that the choice of knot spacing ($\tau_k = 0.5$)
is not creating a significant bias via covariance of the celestial
signal with the low-frequency atmospheric estimate
(see Section~\ref{ssec_bsplines}), we created maps with $\tau_k$ from
$0.15\,\second$ to $1.5\,\second$ for ACT-CL~J0245$-$5301 and
ACT-CL~J0638$-$5358. The temperature profiles for the latter are
plotted in the middle panel of Figure~\ref{fig_profile_s0592}.  Even
the shortest spacing does not produce a profile which is significantly
different from the others. We conclude that the results are not biased
by having knots of too high a frequency.

\begin{\deltab}{llllrrrrrrr}
  \tablecaption{Selection of SZ Clusters Detected by ACT}
  \tablewidth{0pt}
  \tablehead{
    \colhead{ACT Descriptor} & 
    \colhead{Catalog Name} &
    \multicolumn{2}{c}{J2000 Coordinates\tablenotemark{a}} &
    \colhead{rms\tablenotemark{b}} & 
    \colhead{$t_{\mathrm{int}}$\tablenotemark{c}} &
    \colhead{$\Delta T_{\mathrm{SZ}}$\tablenotemark{d}} &
    \multicolumn{3}{c}{$10^{10} \times Y(\theta)$\tablenotemark{e}}\\
    \multicolumn{4}{c}{} &
    \colhead{($\micro\kelvin$)} &
    \colhead{(min)} &
    \colhead{($\micro\kelvin$)} &
    \multicolumn{3}{c}{} \\
    \cline{3-4}\cline{8-10} 
    \multicolumn{2}{c}{} &
    \colhead{\rule{0pt}{2.2ex}RA} &
    \colhead{Dec.} &
    \multicolumn{3}{c}{} &
    \colhead{$\theta \leq 2\arcminute$} &
    \colhead{$\theta \leq 4\arcminute$} &
    \colhead{$\theta \leq 6\arcminute$} \\
    \multicolumn{7}{c}{} &
    \colhead{($\pm \dytwo$)} &
    \colhead{($\pm \dyfour$)} &
    \colhead{($\pm \dysix$)}  
  }
  \startdata
    ACT-CL J0245$-$5301 & AS0295 &
                       \hms{02}{45}{28} & \dms{-53}{01}{36} &
                       44 &
                       10.1 &
                       $-250$ &
                       0.93 & 2.5 & 4.1 \\
    ACT-CL J0330$-$5228 & A3128 (NE) & 
                       \hms{03}{30}{50} & \dms{-52}{28}{38} &
                       49 &
                       10.3 &
                       $-260$ &
                       0.97 & 2.8 & 4.5 \\
    ACT-CL J0546$-$5346 & SPT-CL 0547$-$5345 &
                       \hms{05}{46}{35} & \dms{-53}{46}{04} &
                       46 &
                       9.5 &
                       $-250$ &
                       0.96 & 2.5 & 3.9 \\
    ACT-CL J0638$-$5358 & AS0592 &
                       \hms{06}{38}{46} & \dms{-53}{58}{40} &
                       55 &
                       7.5 &
                       $-230$ &
                       0.74 & 1.5 & 2.2 \\
    ACT-CL J0658$-$5556 & 1E 0657$-$56 (Bullet) & 
                       \hms{06}{58}{33} & \dms{-55}{56}{49} &
                       80 &
                       3.4 &
                       $-510$ &
                       1.70 & 3.20 & 3.80 \\
  \enddata
  \tablenotetext{a}{Position of the deepest point in $2\arcminute$
                    FWHM Gaussian smoothed map.} 
  \tablenotetext{b}{Map rms measured outside a $6\arcminute$ mask and
                    reported for a 1\,arcmin$^2$ area.}
  \tablenotetext{c}{Integration time, defined as the approximate total
                    time (in minutes) that the telescope was pointed
                    in the map region.}
  \tablenotetext{d}{Cluster depth, as measured in a $2\arcminute$ FWHM
                    Gaussian smoothed map at the listed coordinates;
                    intended as a guide to the magnitude of the decrement.} 
  \tablenotetext{e}{See Equation~(\ref{eq_sz_y_int}) and following
                    discussion.} 
  \label{tab_clusters}
\end{\deltab}

\subsection{Integrated Compton-$y$ Values}
\label{ssec_cluster_analysis}

The SZ effect occurs when CMB photons inverse Compton-scatter off
hot electrons in clusters of galaxies
\citep{zeldovich/sunyaev:1969,sunyaev/zeldovich:1970}.
The imprint on the CMB is proportional to the integrated electron gas
pressure:

\begin{equation}
  \frac{\Delta T}{T_{\mathrm{CMB}}} = y\,f(x); \;\;\;\;
  y \equiv \frac{k_B \sigma_T}{m_e c^2} \int dl\,n_e T_e,
  \label{eq_sz}
\end{equation}

\noindent where the integral is along the line of sight, $m_e$, $n_e$,
and $T_e$ are the electron mass, number density, and temperature,
respectively, $\sigma_T$ is the Thomson scattering cross-section, and
the variable $y$ is the Compton-$y$ parameter.  The function
$f(x)$ encodes the dependence on frequency:

\begin{equation}
  f(x) = \left[x \coth(x/2) - 4\right]
         \left[1 + \delta_{\mathrm{SZ}}(x, T_e)\right],
  \label{eq_sz_f}
\end{equation}

\noindent with $x \equiv h\nu / k_B T_{\mathrm{CMB}}$.  The relativistic term
$\delta_{\mathrm{SZ}}$ becomes important at higher
temperatures \citep{rephaeli:1995}, and is taken into account in our
measurements below.

\begin{figure}[htb]
  \centering
\input{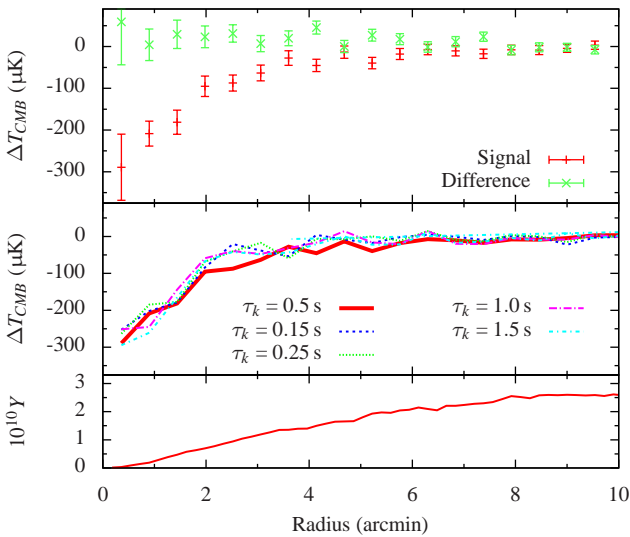}
  \caption{The radial profile (top/middle) and
           integrated Compton $Y(\theta)$ values (bottom) for
           the SZ decrement of ACT-CL~J0638$-$5358 (AS0592).  The
           profile data are averages from the maps in
           $32\arcsecond$~wide annuli, and $Y(\theta)$ is the sum of
           the pixels within a radius $\theta$, converted to the
           unitless Compton-$y$ parameter (Equations~(\ref{eq_sz})
           and (\ref{eq_sz_y_int})).  The top panel shows the profile of the
           signal map and difference map.  The middle
           panel compares profiles for maps made with different knot
           spacings $\tau_k$, showing that the choice of knot spacing
           does not significantly bias the measured cluster profile.
           In all of the panels, the profile centers were determined
           by the minimum of the map after smoothing with a $2'$ FWHM
           Gaussian profile.  (The profiles were calculated from the
           unsmoothed maps.)}    
  \label{fig_profile_s0592}
\end{figure}

A robust measure of the SZ signal is the integrated Compton-$y$
parameter, since it is model-independent and simply sums pixels
in the maps \citep{verde/haiman/spergel:2002,benson/etal:2004}: 

\begin{equation}
  Y(\theta) = \iint\limits_{|\bs{\theta}'| < \theta} 
              d\Omega_{\bs{\theta}'} y(\bs{\theta}'),
  \label{eq_sz_y_int}
\end{equation}

\noindent where $\bs{\theta}$ is the angular distance from the cluster
center.  We use steradians as the unit of solid angle, so $Y$ is
dimensionless.  As an example, it is plotted for ACT-CL~J0638$-$5358 in
the lower panel of Figure~\ref{fig_profile_s0592}.  The values of $Y$ at
$2\arcminute$, $4\arcminute$, and $6\arcminute$ are shown for each
cluster in Table~\ref{tab_clusters}.

For clusters with measured temperatures
(cf. Table~\ref{tab_cluster_param}), our $Y$ values include the
relativistic correction using the formulae from \citet{nozawa/etal:2006}.  At
$148\,\giga\hertz$, the corrections increase $Y$ and, for the clusters
in this paper, range from 4\% (ACT-CL~J0330$-$5228) to 7\%
(ACT-CL~J0658$-$5556).   Thus, the $Y$ value quoted for
ACT-CL~J0546$-$5346, for which there is no measured temperature, is
biased low, though we note that the relativistic corrections for the
other clusters are smaller than the uncertainty of our measurements
(see below).

There are two statistical sources of uncertainty in the measured $Y$ values:
instrumental and atmospheric noise in the map, and confusion of
primary CMB anisotropies with the SZ signal.  The contribution from
noise is readily estimated from the map rms
(cf. Sec.~\ref{ssec_cluster_data} and Table~\ref{tab_clusters}).  
To estimate the second contribution, we found the standard deviation
of $Y$ in the ensemble of simulated, noise-free, primary CMB-only maps
described in Section~\ref{ssec_profiles}.  This latter source dominates over
the error from the map noise.  Our estimated $1\sigma$ errors,
including both sources of uncertainty, are $\dytwo\e{-10}$,
$\dyfour\e{-10}$ and $\dysix\e{-10}$ for $Y$ at $2\arcminute$,
$4\arcminute$, and $6\arcminute$.  We arrive at uncertainties that
agree to better than 10\% by calculating the standard deviation of $Y$
values from the 24 maps of random patches of the sky with no clusters
present (cf. Section~\ref{ssec_profiles}).

Uncertainty in the cluster temperature enters into the measurement of
$Y$ via the relativistic correction (see above).  However, the error
introduced is less than a percent and is therefore insignificant in
comparison to the contributions from noise and primary CMB anisotropies.

\subsection{Comparisons to Previous Measurements}
\label{ssec_cluster_compare}

The clusters shown in this paper are previously known X-ray, optical,
and/or SZ clusters; all are massive systems.  
For three of the sources
(AS0295, A3128 (NE), and AS0592), 
these are
the first reported SZ detections.  In this section, we briefly review
measurements from the literature to provide context, and point out
some of the contributions that our new measurements make to this body
of knowledge. 

Relevant parameters from the literature are listed in
Table~\ref{tab_cluster_param}; references for these values are
included below. Typical errors on $L_X$ are small ($<$ 20\%), while
those on the inferred mass are more substantial ($\sim$50\%).
Temperatures are measured values from X-ray spectra.  We use a flat
$\Lambda$CDM cosmology with $\Omega_M=0.3$ and 
$H_0 = 70\,\kilo\metre\,\second^{-1}\,\mathrm{Mpc}^{-1}$.  Masses are
quoted in units of $M_{500}$, defined as the mass 
within a radius having a mean mass density $\langle\rho\rangle$ 500 times
greater than the critical density, i.e., 
$\langle\rho\rangle = 500\times3H^2/(8\pi G)$.  In the following we
briefly discuss the clusters in the order in which they appear in
Table~\ref{tab_clusters}.

\begin{\deltab}{llrccccc}
  \tablecaption{Summary of Cluster Properties from X-Ray and Optical Studies}
  \tablewidth{0pt}
  \tablehead{
    \colhead{ACT Descriptor} &
    \colhead{Catalog Name} &
    \colhead{Redshift} &
    \colhead{$D_{\rm A}$} &
    \colhead{$L_{\rm X}$(0.1--2.4~keV)} &
    \colhead{$M_{500}$} &
    \colhead{$kT$} &
    \colhead{$10^{10} \times Y_{2500}$\tablenotemark{a}} \\
    \multicolumn{3}{c}{} &
    \multicolumn{1}{c}{[Mpc]} &
    \multicolumn{1}{c}{($10^{44}$~erg~s$^{-1}$)}  &
    \multicolumn{1}{c}{($10^{15}M_{\sun}$)} &
    \multicolumn{1}{c}{(keV)} &
    \multicolumn{1}{c}{} 
  }
  \startdata
    ACT-CL J0245$-$5301 &  AS0295     &  $0.3006$   & $\phantom{1}920$ 
                     & $\phantom{1}8.3$ & $0.8$ & $\phantom{1}6.7\pm0.7$ 
                     & $0.53^{+0.35}_{-0.21}$ \\

    ACT-CL J0330$-$5228 &  A3128 (NE) &  $0.44$     & $1172$           
                     & $\phantom{1}3.9$ & $0.3$ & $\phantom{1}5.1\pm0.2$ 
                     & $0.15^{+0.10}_{-0.06}$ \\

    ACT-CL J0546$-$5346 & SPT~0547$-$5345  &  $0.88$ (P) & $1596$           
                     & $\phantom{1}4.7$ & $0.6$  & --- 
                     & --- \\

    ACT-CL J0638$-$5358 &  AS0592     &  $0.2216$   & $\phantom{1}737$ 
                     & $10.6$           & $1.0$  & $\phantom{1}8.0\pm0.4$ 
                     & $1.31^{+0.86}_{-0.52}$ \\

    ACT-CL J0658$-$5556 &  1E 0657$-$56    &  $0.296$    & $\phantom{1}910$ 
                     & $20.5$           & $1.4$ & $10.6\pm0.1$  
                     & $1.61^{+1.16}_{-0.67}$ \\
  \enddata
  \tablecomments{See Section~\ref{ssec_cluster_compare} for citations
                 to the literature from which these values were obtained.
                 The marker (P) in the redshift column indicates a
                 photometric redshift measurement.}
  \tablenotetext{a}{Predicted value of $Y$ within $R_{2500}$ from the
                    $Y$-$kT$ scaling relation of
                    \citet{bonamente/etal:2008}.  Errors come from 
                    the uncertainty on the scaling relation
                    parameters.  Although we do not have $R_{2500}$
                    values for our clusters, the $Y(2\arcminute)$
                    measurements listed in Table~\ref{tab_clusters} should be
                    roughly comparable to these---see
                    Section~\ref{sssec_sz_compare}.} 
  \label{tab_cluster_param}
\end{\deltab}

\subsubsection{AS0295}

AS0295 first appeared in \citet{abell/corwin/olowin:1989} in
their table of supplementary southern clusters (i.e., clusters that
were not rich enough or were too distant to satisfy the criteria for
inclusion in the rich nearby cluster catalog).  It was also found to
be a significant X-ray source in the {\em ROSAT} All Sky Survey (RASS)
Bright Source Catalog \citep{voges/etal:1999}.  The spectroscopic
redshift of AS0295 was obtained by \citet{edge/etal:1994}, who
also reported the discovery of a giant strong-lensing arc near the
brightest cluster galaxy. Efforts to detect the SZ effect at 1.2\,mm
and 2\,mm with the Swedish-ESO Submillimetre Telescope
were attempted, unsuccessfully, by \citet{andreani/etal:1996}.  {\it
ASCA} observations \citep{fukazawa/makishima/ohashi:2004} yielded
values (see Table~\ref{tab_cluster_param}) for average temperature and
soft band X-ray flux(0.1--2.4 keV), from which we determined the
corresponding X-ray luminosity.  The cluster mass $M_{500}$ was then
estimated from the luminosity-mass (specifically $L_X$(0.1-2.4~keV)
versus $M_{500}$) relations from \citet{reiprich/bohringer:2002}.

\subsubsection{A3128 (NE)}

\begin{figure}[tb]
  \centering
\input{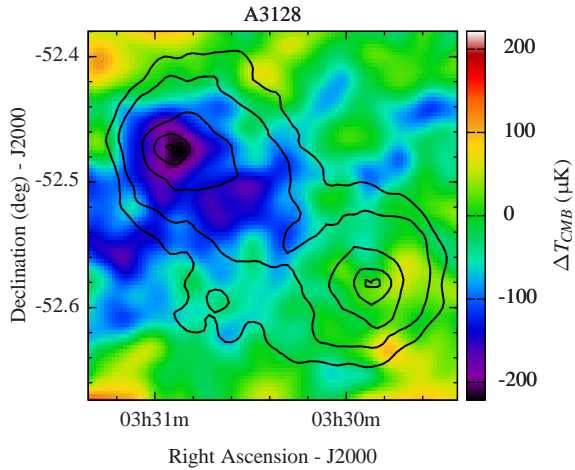}
  \caption{ACT-CL~J0330-5228 (A3128 (NE)) with overlaid contours
           of X-ray emission in black.  The SZ detection is associated
           with the NE feature of A3128, and confirms that it is
           due to a more massive, higher redshift cluster than that at
           the SW lobe---a compelling example of the redshift
           independent mass selection of the SZ effect.  The X-ray data
           come from two separate {\em XMM-Newton} observations (Obs Ids
           0400130101 and 0400130201) with a total exposure time of
           104\,ks.  The two observations were mosaicked into a single 
           image over the 0.2--2.0 keV.  Contour values are from
           $1.25\e{-8}$ to $1.25\e{-7}$ 
           photons\,cm$^{-2}$\,s$^{-1}$\,arcsec$^{-2}$.} 
  \label{fig_a3128_overlay}
\end{figure}

Until quite recently the north-east (NE) component of A3128 was
believed to be part of the Horologium-Reticulum supercluster at
$z=0.06$.  The X-ray morphology is clearly double peaked with the two
components separated on the sky by some
$12\arcminute$.  \citet{rose/etal:2002} estimated the virial masses of
the two components assuming the redshift of the supercluster and
obtained a value for each of $\sim$$1.5\times10^{14}\, M_\odot$.
Figure~\ref{fig_a3128_overlay} shows our SZ measurement with overlaid X-ray
contours.

Recently \citet{werner/etal:2007} carried out a detailed study of
this cluster using {\em XMM-Newton} data, which revealed a more distant and
more massive cluster superposed on the northeastern component of
A3128.  A significant portion of the X-ray emission comes from this 
background cluster.  The values we quote in the table for redshift,
X-ray luminosity, gas temperature, and $M_{500}$ correspond to the
background cluster and come from \citet{werner/etal:2007}.

The large SZ decrement seen in the ACT maps is clearly
associated with the NE component where the $z=0.44$ cluster is.  We do
not detect a significant decrement from the southwestern component
which lies at $z=0.06$. \citet{werner/etal:2007} estimate the
temperature of the higher redshift cluster to be $5.14 \pm 0.15$ keV,
which is significantly hotter than that of the foreground cluster
($kT=3.36\pm0.04$ keV).  This system, therefore, is a compelling
illustration of the mass selection, approximately independent of
redshift, of the SZ effect.  \citet{werner/etal:2007} note that the
temperature, luminosity, and mass estimates of the $z=0.44$ background
cluster are all subject to large systematic errors, as the cluster
properties depend upon the assumed properties of the foreground
system.  A joint X-ray/SZ/optical analysis should be able to better
constrain the characteristics of both systems and thereby contribute
to assessing the mass threshold of the ACT cluster survey.

\subsubsection{SPT 0547$-$5345}

The galaxy cluster SPT~0547$-$5345 was first discovered via the SZ
effect by the South Pole
Telescope \citep{staniszewski/etal:2009}. Its
physical properties---photometric redshift, luminosity, temperature
and mass estimate---were subsequently reported
by \citet{menanteau/hughes:2009} based on optical and X-ray data.

SPT~0547$-$5345 has associated central elliptical galaxies with
luminosities consistent with those of clusters in the Sloan Digital
Sky Survey.  Its mass estimate from the optical and X-ray luminosity
also suggests that it is a fairly massive system.  In
Table~\ref{tab_cluster_param} we list its $M_{500}$ estimates from
the X-ray luminosity.

\subsubsection{AS0592}

The galaxy cluster AS0592 was originally detected optically
\citep{abell/corwin/olowin:1989}.  {\em ROSAT} detected it as a bright
source in the All Sky Survey  and its
redshift ($z=0.2216$) was reported in \citet{degrandi/etal:1999}. The
cluster is also known by its REFLEX designation of RXC J0638.7$-$5358
\citep{bohringer/etal:2004}.  The {\em ROSAT} flux and luminosity in
the soft X-ray band (0.1-2.4 keV) are $7.5\times 10^{-12}$ erg
cm$^{-2}$ s$^{-1}$ and $1.1\times 10^{45}$ ergs s$^{-1}$. The X-ray
spectrum of AS0592 from a {\em Chandra} observation
\citep{hughes/etal:2009} yields an integrated gas temperature of
$kT=8.0\pm0.4$ keV. The soft X-ray luminosity implies a cluster mass
of $M_{500} = 10^{15}\,M_\odot$.

\subsubsection{1E 0657$-$56 (Bullet Cluster)}

We detect 1E 0657$-$56, the famous ``Bullet'' cluster, at high
significance with a strong central decrement and large integrated $Y$.
Previous detections of the millimeter-band SZ signal from this cluster
have been reported by ACBAR \citep{gomez/etal:2004} and
APEX-SZ \citep{halverson/etal:2008}.

The spectroscopic redshift of 1E 0657$-$56 was obtained by
\citet{tucker/etal:1998}, the X-ray flux came from the Einstein
Observatory \citep{markevitch/etal:2002}, the X-ray gas temperature
from {\em XMM-Newton} \citep{zhang/etal:2006}, and the cluster mass,
$M_{500}$, from a study by \citet{zhang/etal:2008}.

\begin{figure}[tb]
  \centering
\input{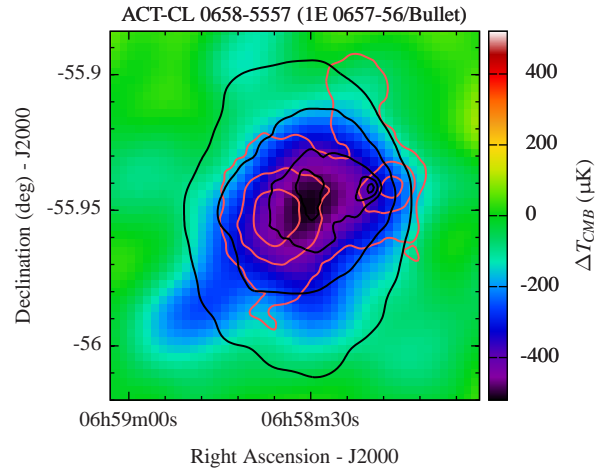}
  \caption{ACT-CL~J0658-5556 (Bullet Cluster) with overlaid contours
           of X-ray emission (black) and dark matter distribution
           (orange).  The X-ray contours come from an
           $85\,\kilo\second$-long {\em Chandra} observation (Obs Id
           3184) and correspond to the 0.5-2.0 keV band.  Contour
           values are $4\e{-7}$ to $2\e{-9}$
           photons\,cm$^{-2}$\,s$^{-1}$\,arcsec$^{-2}$.  The lensing
           data are from \citet{clowe/randall/markevitch:2007} with
           contours running from $\kappa = 0.12$ to $0.39$.}
  \label{fig_bull_overlay}
\end{figure}

Figure~\ref{fig_bull_overlay} shows a zoomed-in plot of our SZ map with
X-ray contours from {\em Chandra} and lensing contours from
\citet{clowe/randall/markevitch:2007}.  As expected, the SZ decrement
follows the X-ray contours more closely than the lensing contours,
since the collisionless dark matter is expected to be offset from the
collisional gas in this merging system.

\subsubsection{Comparison with Previous SZ Measurements}
\label{sssec_sz_compare}

Although the large masses of the ACT-detected clusters we report
here offer strong support for the reality of our detections, we also
compare the quoted integrated Compton-$y$ parameters for consistency
with expectations from previous SZ cluster studies.  For this we use
the $Y$-$kT$ 
scaling relation from \citet{bonamente/etal:2008} (using values for
``all clusters'' from their Table 2).  Predicted values are given in the last
column of Table~\ref{tab_cluster_param}.   The $Y$ values in the scaling
relation were integrated within $R_{2500}$, the radius where the average cluster
mass density is 2500 times the critical density.  We do not have precise
$R_{2500}$ values for our clusters, but estimates of $R_{2500}$ range from about
$1\arcminute$ to $3\arcminute$, so the predicted values of $Y(2500)$ should,
to first order, be roughly comparable to our $Y(2\arcminute)$ values.  
With that proviso, the predicted and measured $Y$ values agree to
within $2\sigma$ for all clusters except A3128 (NE), where the
cluster temperature predicts a much lower $Y$ value than we measure.
Since it is a complex system, it could have a larger mass than
previously thought.  A larger and better-studied sample of
ACT-detected clusters will be necessary before drawing conclusions
about scaling relations.

\section{Conclusions}
\label{sec_conclusions}

In this paper we have presented beam maps and SZ cluster measurements
from the first full season of \arone\ data from ACT.  They
represent the first scientific results from ACT and demonstrate that
it is poised to make important contributions to millimeter astronomy.

We have described a maximum-likelihood mapping algorithm which uses
B-splines to model atmospheric signal and to remove it from the data.
The method has been used to make high precision ($<-40\,\deci\bel$)
beam maps, with solid angles in the \arone, \artwo, and \arthree\ bands
of $(218.2 \pm 4)\,\nsr$, $(118.2 \pm 3)\,\nsr$, and $(104.2 \pm
6)\,\nsr$, respectively.  The beam profiles and window functions will
be important for all subsequent analyses of ACT's data.

Additionally, we have made maps showing SZ decrements of five
previously discovered galaxy clusters.  Of these, three are detected
for the first time via the SZ effect. Our high-significance detection of 
the $z = 0.44$ component of A3128, and our current
non-detection of the low-redshift part, corroborates existing
evidence that the further cluster is more massive.  This is a
compelling example of the redshift-independent mass selection of the
SZ effect which will be exploited in future studies of ACT clusters.

A preprint of this paper included three additional clusters:
ACT-CL~J0509$-$5345, ACT-CL~J0516$-$5432, and ACT-CL J0645$-$5413.
They are instead quantified in \citet{menanteau/etal:2010}
and \citet{marriage/etal:prepb}.  We omitted them here in order to
focus on the more significant of the initial ACT cluster detections.
\\
\\
The ACT project was proposed in 2000 and
funded 2004 Jan 1. Many have contributed to the project since its
inception. We especially wish to thank Asad Aboobaker, Christine
Allen, Dominic Benford, Paul Bode, Kristen Burgess, Angelica
de Oliveria-Costa, Peter Hargrave, Norm Jarosik, Amber Miller, Carl
Reintsema, Uros Seljak, Martin Spergel, Johannes Staghun, Carl Stahle,
Max Tegmark, Masao Uehara, and Ed Wishnow. It is a pleasure to
acknowledge Bob Margolis, ACT's project manager. Reed Plimpton and
David Jacobson worked at the telescope during the 2008 season. ACT is
on the Chajnantor Science preserve which was made possible by CONICYT.  We
are grateful for the assistance we received at various times  from the
ALMA, APEX, ASTE, CBI/QUIET, and NANTEN2 groups.  The PWV data come
from the public APEX weather site.  Field operations were based at the
Don Esteban facility run by Astro-Norte.  This research has made use
of the NASA/IPAC Extragalactic Database (NED) which is operated by the
Jet Propulsion Laboratory, California Institute of Technology, under
contract with the National Aeronautics and Space Administration.
Satoshi Nozawa and Naoki Itoh kindly shared their code for calculating
relativistic corrections to the SZ effect.  We also thank William Holzapfel
and an anonymous referee who provided helpful feedback on an earlier
version of this paper. We thank the members of our external advisory
board---Tom Herbig (chair), Charles Alcock, Walter Gear, Cliff
Jackson, Amy Newbury, and Paul Steinhardt---who helped guide the
project to fruition.    This work was supported by the U.S. National
Science Foundation through awards AST-0408698 for the ACT project, and
PHY-0355328, AST-0707731 and PIRE-0507768. Funding was also provided
by Princeton University and the University of Pennsylvania. A.D.H.
received additional support from a Natural Science and Engineering
Research Council of Canada (NSERC) PGS-D scholarship. A.K. and B.P. were
partially supported through NSF AST-0546035 and AST-0606975,
respectively, for work on ACT.  H.Q. and L.I. acknowledge partial support
from FONDAP Centro de Astrofisica.  E.S. acknowledges support by NSF
Physics Frontier Center grant PHY-0114422 to the Kavli Institute of
Cosmological Physics. K.M., M.H., and R.W. received financial support
from the South African National Research Foundation (NRF), the Meraka
Institute via funding for the South African Centre for High
Performance Computing (CHPC), and the South African Square Kilometer
Array (SKA) Project.  R.H. received funding from the Rhodes Trust.
L.V. acknowledges support from NSF-AST 0707731 and FP7-PEOPLE-2007-4-3
IRG no. 202182.  J.B.J. acknowledges for support from FONDECYT (no:
3085031).

\appendix

\section{The Cottingham Method as a Maximum-Likelihood Estimator}
\label{appendix_mle}

Equation~(\ref{eq_signal_model}) can be written in the matrix form,

\begin{equation}
  \bvec{d} = \left(\begin{array}{cc}\bm{P} & \bm{B}\end{array}\right) 
           \left(\begin{array}{c} \bvec{m} \\
                                  \bs{\alpha}\end{array}\right) 
           + \bvec{n},
  \label{eq_sigmal_model_matrix}
\end{equation}

\noindent which has the maximum-likelihood estimator:

\begin{equation}
  \left(\begin{array}{cc}\wt{\bvec{m}} \\
                         \wt{\bs{\alpha}}\end{array}\right) 
   = \left[\left(\begin{array}{c} \bm{P}^T \\
                                  \bm{B}^T\end{array}\right) 
     \bm{N}^{-1}
     \left(\begin{array}{cc} \bm{P} &
                             \bm{B}\end{array}\right)\right]^{-1}
     \left(\begin{array}{c} \bm{P}^T \\ \bm{B}^T\end{array}\right)
     \bm{N}^{-1}\bvec{d}
   = \left[\begin{array}{cc} (\bm{P}\bm{P}) & (\bm{P}\bm{B}) \\
           (\bm{B}\bm{P}) & (\bm{B}\bm{B})\end{array}\right]^{-1}
     \left(\begin{array}{c} \bm{P}^T\bm{N}^{-1}\bvec{d} \\
                            \bm{B}^T\bm{N}^{-1}\bvec{d}\end{array}\right),
  \label{eq_mle_matrix}
\end{equation}

\noindent where we use the shorthand notation 
$(\bm{X}\bm{Y}) \equiv \bm{X}^T\bm{N}^{-1}\bm{Y}$.  The inverted
matrix evaluates to:

\begin{equation}
  \left[\begin{array}{cc}
    \left[(\bm{PP}) - (\bm{PB})(\bm{BB})^{-1}(\bm{BP})\right]^{-1} &
      -(\bm{PP})^{-1}(\bm{PB})\left[(\bm{BB}) -
       (\bm{BP})(\bm{PP})^{-1}(\bm{PB})\right]^{-1} \\
    -(\bm{BB})^{-1}(\bm{BP})\left[(\bm{PP}) -
      (\bm{PB})(\bm{BB})^{-1}(\bm{BP})\right]^{-1} &
      \left[(\bm{BB}) - (\bm{BP})(\bm{PP})^{-1}(\bm{PB})\right]^{-1}
  \end{array}\right].
  \label{eq_mle_matrix_inverse}
\end{equation}

\noindent Since this matrix is symmetric, and $(\bm{XY})^T =
(\bm{YX})$, we can rewrite the lower-left component as:

\begin{align}
  (\bm{BB})^{-1}(\bm{BP})\left[(\bm{PP}) -
  (\bm{PB})(\bm{BB})^{-1}(\bm{BP}) \right]^{-1}
  &= \left\lbrace(\bm{PP})^{-1}(\bm{PB})\left[(\bm{BB}) -
     (\bm{BP})(\bm{PP})^{-1}(\bm{PB})\right]^{-1}\right\rbrace^T \noindent\\
  &= \left[(\bm{BB}) - (\bm{BP})(\bm{PP})^{-1}(\bm{PB})\right]^{-1}
     (\bm{BP})(\bm{PP})^{-1}.
\end{align}

\noindent Thus, the solution for the atmosphere is:

\begin{align}
  \wt{\bs{\alpha}} \;=\; & \left[(\bm{BB})- 
  (\bm{BP})(\bm{PP})^{-1}(\bm{PB})\right]^{-1}
  \bm{B}^T\bm{N}^{-1}\bvec{d} \;\;-\;\;
  \left[(\bm{BB}) - (\bm{BP})(\bm{PP})^{-1}(\bm{PB})\right]^{-1}
  (\bm{BP})(\bm{PP})^{-1}\bm{P}^T\bm{N}^{-1}\bvec{d}.
  \label{eq_mle_alpha}
\end{align}

\noindent To show that this is equivalent to the solution presented in
Section~\ref{ssec_cott_algorithm}, we observe that the definitions in 
Equations~(\ref{eq_def_proj}) and (\ref{eq_def_xi_theta_phi}) of
Section~\ref{ssec_cott_algorithm} can be recast: 

\begin{align}
  \bs{\Theta} &\equiv \bm{B}^T\bm{N}^{-1}(\bs{1}-\bm{P}\bs{\Pi})\bm{B} \;\;=\;\;
    (\bm{BB}) - (\bm{BP})(\bm{PP})^{-1}(\bm{PB}), \nonumber\\
  \bs{\phi} &\equiv \bm{B}^T\bm{N}^{-1}(\bs{1}-\bm{P}\bs{\Pi})\bvec{d} \;\;=\;\;
    \left[\bm{B}^T\bm{N}^{-1} - (\bm{BP})(\bm{PP})^{-1}\bm{P}^T\bm{N}^{-1}
    \right]\bvec{d}.
  \label{eq_def_theta_phi_recast}
\end{align}

\noindent This reduces Equation~(\ref{eq_mle_alpha}) to:

\begin{equation}
  \wt{\bs{\alpha}} = \bs{\Theta}^{-1}\bs{\phi},
\end{equation}

\noindent which is the same as Equation~(\ref{eq_cott_linear_eq}) of
Section~\ref{ssec_cott_algorithm}.

\section{A Planet's Solid Angle Contribution to the Beam Solid Angle
         Measurement}  
\label{appendix_sa}

Denote the instrument response with $P(\bvec{n})$ and the power emitted
by the planet with $P_0\Psi(\bvec{n})$, where $P_0$ is the peak power
emitted and $\Psi$ is a normalized distribution describing its shape.
The coordinate $\bvec{n}$ is a two dimensional vector describing the
position on the sky, with $\bvec{n} = \bvec{0}$ at the planet center.  The
measured beam map, $\wt{B}$, is the convolution of the true beam, $B$,
with the planet:

\begin{equation}
  \wt{B}(\bvec{n}) = \frac{P(\bvec{n})}{P(\bvec{0})}
  = \frac{\iint d\Omega_{\bvec{n}'} B(\bvec{n}-\bvec{n}')\Psi(\bvec{n}')}
         {\iint d\Omega_{\bvec{n}'} B(-\bvec{n}')\Psi(\bvec{n}')}.
  \label{eq_beam_convolution}
\end{equation}

\noindent The measured solid angle is then
(cf. Equation~(\ref{eq_def_accum_sa})):

\begin{equation}
  \wt{\Omega}_A = \iint d\Omega_{\bvec{n}}
    \frac{\iint d\Omega_{\bvec{n}'} B(\bvec{n}-\bvec{n}')\Psi(\bvec{n}')}
    {\iint d\Omega_{\bvec{n}'} B(-\bvec{n}')\Psi(\bvec{n}')} =
    \frac{\left[\iint d\Omega_{\bvec{n}'} \Psi(\bvec{n}')\right]
          \left[\iint d\Omega_{\bvec{n}} B(\bvec{n})\right]}
         {\iint d\Omega_{\bvec{n}'} B(\bvec{n}')\Psi(\bvec{n}')} =
    \frac{\Omega_{\Psi}\Omega_A}
         {\iint d\Omega_{\bvec{n}'} B(\bvec{n}')\Psi(\bvec{n}')},
  \label{eq_sa_convolution}
\end{equation}

\noindent where in the second equality we brought the denominator
outside the outer integral, and in the numerator we switched the order of
integration and then shifted the dummy variable for the integral over
$B$.  In the last equality we recognized that the integrals in the
numerator evaluate to the solid angles of the planet and the true
instrument beam, respectively.  If the planet is much smaller than the
beam, we can expand the beam appearing the integrand of the
denominator in a Taylor series:

\begin{equation}
  B(\bvec{n}) = 1 \;+\; \nabla B(\bvec{0}) \cdot \bvec{n} \;+\; 
              \frac{1}{2} \bvec{n}\cdot\bm{H}(\bvec{0})\cdot\bvec{n} \;+\;
              \cdots, 
  \label{eq_beam_taylor}
\end{equation}

\noindent where $\bm{H}$ is the Hessian matrix of the beam.  At the beam
center, being the peak, the gradient vanishes.  If we assume a
symmetric beam, then
$\bvec{n}\cdot\bm{H}\cdot\bvec{n} = (\nabla^2 B / 2)|\bvec{n}|^2$, and we can
write: 

\begin{equation}
  \wt{\Omega}_A \approx \frac{\Omega_{\Psi}\Omega_A}
  {\iint d\Omega_{\bvec{n}'}\left[1 + \frac{1}{4} 
         \nabla^2 B(\bvec{0})|\bvec{n}'|^2\right]\Psi(\bvec{n}')} = 
  \Omega_A \left[1 + \frac{\nabla^2 B(\bvec{0})}{4\Omega_{\Psi}}
                 \iint d\Omega_{\bvec{n}'}|\bvec{n}'|^2
                       \Psi(\bvec{n}')\right]^{-1} =
  \Omega_A \left[1 + \frac{\nabla^2B(\bvec{0})}{4\Omega_{\Psi}}
                     \mu^{\Psi}_2\right]^{-1},
  \label{eq_sa_conv_approx_1}
\end{equation}

\noindent where $\mu^{\Psi}_2$ is the second raw moment of the planet
shape $\Psi$.  In the small planet approximation we are making, the
second term in the brackets is small.  Thus:

\begin{equation}
  \wt{\Omega}_A \approx \Omega_A - \frac{\Omega_A}{\Omega_{\Psi}}
                \frac{\nabla^2B(\bvec{0})}{4}\mu^{\Psi}_2.
\end{equation}

\noindent For a disk, $\mu^{\Psi}_2 = \Omega_{\Psi}^2 / 2\pi$, and both a
Gaussian beam and an Airy pattern have $\nabla^2 B(\bvec{0}) =
-4\pi/\Omega_A$.  (For the Airy pattern, this is easiest to see
by expanding the Bessel function in a power series and
differentiating.) Thus, we have the result that
$\wt{\Omega}_A \approx \Omega_A + \Omega_{\Psi}/2$.

\bibliography{act}

\begin{thebibliography}{72}
\expandafter\ifx\csname natexlab\endcsname\relax\def\natexlab#1{#1}\fi

\bibitem[{{Abell} {et~al.}(1989){Abell}, {Corwin}, \&
  {Olowin}}]{abell/corwin/olowin:1989}
{Abell}, G.~O., {Corwin}, Jr., H.~G., \& {Olowin}, R.~P. 1989, \apjs, 70, 1

\bibitem[{{Andreani} {et~al.}(1996){Andreani}, {Pizzo}, {dall'Oglio},
  {Whyborn}, {Boehringer}, {Shaver}, {Lemke}, {Otarola}, {Nyman}, \&
  {Booth}}]{andreani/etal:1996}
{Andreani}, P. {et~al.} 1996, \apjl, 459, L49, arXiv:astro-ph/9602011

\bibitem[{Battistelli {et~al.}(2008)Battistelli, Amiri, Burger, Devlin, Dicker,
  Doriese, D{\"{u}}nner, Fisher, Fowler, Halpern, Hasselfield, Hilton, Hincks,
  Irwin, Kaul, Klein, Knotek, Lau, Limon, Marriage, Niemack, Page, Reintsema,
  Staggs, Swetz, Switzer, Thornton, \& Zhao}]{battistelli/etal:2008}
Battistelli, E.~S. {et~al.} 2008, in Proc.\ SPIE, ed. W.~D. Duncan, W.~S.
  Holland, S.~Withington, \& J.~Zmuidzinas, Vol. 7020 (SPIE), 702028

\bibitem[{{Benson} {et~al.}(2004){Benson}, {Church}, {Ade}, {Bock}, {Ganga},
  {Henson}, \& {Thompson}}]{benson/etal:2004}
{Benson}, B.~A., {Church}, S.~E., {Ade}, P.~A.~R., {Bock}, J.~J., {Ganga},
  K.~M., {Henson}, C.~N., \& {Thompson}, K.~L. 2004, \apj, 617, 829,
  arXiv:astro-ph/0404391

\bibitem[{{B{\"o}hringer} {et~al.}(2004){B{\"o}hringer}, {Schuecker}, {Guzzo},
  {Collins}, {Voges}, {Cruddace}, {Ortiz-Gil}, {Chincarini}, {De Grandi},
  {Edge}, {MacGillivray}, {Neumann}, {Schindler}, \&
  {Shaver}}]{bohringer/etal:2004}
{B{\"o}hringer}, H. {et~al.} 2004, \aap, 425, 367, arXiv:astro-ph/0405546

\bibitem[{{Bojanov} {et~al.}(1993){Bojanov}, {Hakopian}, \&
  {Sahakian}}]{bojanov/hakopian/sahakian:SFAMI}
{Bojanov}, B.~D., {Hakopian}, H.~A., \& {Sahakian}, A.~A. 1993, {Spline
  Functions and Multivariate Interpolations} ({Kluwer Academic Publishers})

\bibitem[{{Bonamente} {et~al.}(2008){Bonamente}, {Joy}, {LaRoque}, {Carlstrom},
  {Nagai}, \& {Marrone}}]{bonamente/etal:2008}
{Bonamente}, M., {Joy}, M., {LaRoque}, S.~J., {Carlstrom}, J.~E., {Nagai}, D.,
  \& {Marrone}, D.~P. 2008, \apj, 675, 106, 0708.0815

\bibitem[{{Bond}(1996)}]{bond:1996}
{Bond}, J.~R. 1996, in Cosmology and Large Scale Structure, ed. R.~{Schaeffer},
  J.~{Silk}, M.~{Spiro}, \& J.~{Zinn-Justin}, 469

\bibitem[{Born \& Wolf(1999)}]{born/wolf:POO:7e}
Born, M., \& Wolf, E. 1999, {Principles of Optics}, 7th edn. ({Cambridge
  University Press})

\bibitem[{{Boughn} {et~al.}(1992){Boughn}, {Cheng}, {Cottingham}, \&
  {Fixsen}}]{boughn/etal:1992}
{Boughn}, S.~P., {Cheng}, E.~S., {Cottingham}, D.~A., \& {Fixsen}, D.~J. 1992,
  \apjl, 391, L49

\bibitem[{{Burigana} {et~al.}(1999){Burigana}, {Malaspina}, {Mandolesi},
  {Danse}, {Maino}, {Bersanelli}, \& {Maltoni}}]{burigana/etal:1999}
{Burigana}, C., {Malaspina}, M., {Mandolesi}, N., {Danse}, L., {Maino}, D.,
  {Bersanelli}, M., \& {Maltoni}, M. 1999, Internal Report ITESRE 198/1997,
  arXiv:astro-ph/9906360

\bibitem[{{Clowe} {et~al.}(2007){Clowe}, {Randall}, \&
  {Markevitch}}]{clowe/randall/markevitch:2007}
{Clowe}, D., {Randall}, S.~W., \& {Markevitch}, M. 2007, Nuclear Physics B
  Proceedings Supplements, 173, 28, arXiv:astro-ph/0611496

\bibitem[{{Cottingham}(1987)}]{cottingham:1987}
{Cottingham}, D.~A. 1987, PhD thesis, Princeton University

\bibitem[{{de Boor}(2001)}]{deboor:APGTS:rev}
{de Boor}, C. 2001, {A Practical Guide to Splines}, {revised} edn. ({Springer})

\bibitem[{{de Grandi} {et~al.}(1999){de Grandi}, {B{\"o}hringer}, {Guzzo},
  {Molendi}, {Chincarini}, {Collins}, {Cruddace}, {Neumann}, {Schindler},
  {Schuecker}, \& {Voges}}]{degrandi/etal:1999}
{de Grandi}, S. {et~al.} 1999, \apj, 514, 148, arXiv:astro-ph/9902067

\bibitem[{{Delabrouille}(1998)}]{delabrouille:1998}
{Delabrouille}, J. 1998, \aaps, 127, 555

\bibitem[{{Edge} {et~al.}(1994){Edge}, {Boehringer}, {Guzzo}, {Collins},
  {Neumann}, {Chincarini}, {de Grandi}, {Duemmler}, {Ebeling}, {Schindler},
  {Seitter}, {Vettolani}, {Briel}, {Cruddace}, {Gruber}, {Gursky}, {Hartner},
  {MacGillivray}, {Schuecker}, {Shaver}, {Voges}, {Wallin}, {Wolter}, \&
  {Zamorani}}]{edge/etal:1994}
{Edge}, A.~C. {et~al.} 1994, \aap, 289, L34, arXiv:astro-ph/9407078

\bibitem[{Fowler {et~al.}(2007)Fowler, Niemack, Dicker, Aboobaker, Ade,
  Battistelli, Devlin, Fisher, Halpern, Hargrave, Hincks, Kaul, Klein, Lau,
  Limon, Marriage, Mauskopf, Page, Staggs, Swetz, Switzer, Thornton, \&
  Tucker}]{fowler/etal:2007}
Fowler, J.~W. {et~al.} 2007, Appl. Opt., 46, 3444

\bibitem[{{Friedman} {et~al.}(2009){Friedman}, {Ade}, {Bock}, {Bowden},
  {Brown}, {Cahill}, {Castro}, {Church}, {Culverhouse}, {Ganga}, {Gear},
  {Gupta}, {Hinderks}, {Kovac}, {Lange}, {Leitch}, {Melhuish}, {Memari},
  {Murphy}, {Orlando}, {O'Sullivan}, {Piccirillo}, {Pryke}, {Rajguru},
  {Rusholme}, {Schwarz}, {Taylor}, {Thompson}, {Turner}, {Wu}, {Zemcov}, \&
  {QUa D collaboration}}]{friedman/etal:2009}
{Friedman}, R.~B. {et~al.} 2009, \apjl, 700, L187, 0901.4334

\bibitem[{{Fukazawa} {et~al.}(2004){Fukazawa}, {Makishima}, \&
  {Ohashi}}]{fukazawa/makishima/ohashi:2004}
{Fukazawa}, Y., {Makishima}, K., \& {Ohashi}, T. 2004, \pasj, 56, 965,
  arXiv:astro-ph/0411745

\bibitem[{{Ganga} {et~al.}(1993){Ganga}, {Cheng}, {Meyer}, \&
  {Page}}]{ganga/etal:1993}
{Ganga}, K., {Cheng}, E., {Meyer}, S., \& {Page}, L. 1993, \apjl, 410, L57

\bibitem[{{Gomez} {et~al.}(2004){Gomez}, {Romer}, {Peterson}, {Chase},
  {Runyan}, {Holzapfel}, {Kuo}, {Newcomb}, {Ruhl}, {Goldstein}, \&
  {Lange}}]{gomez/etal:2004}
{Gomez}, P. {et~al.} 2004, in American Institute of Physics Conference Series,
  Vol. 703, Plasmas in the Laboratory and in the Universe: New Insights and New
  Challenges, ed. G.~{Bertin}, D.~{Farina}, \& R.~{Pozzoli}, 361--366

\bibitem[{{Griffin} \& {Orton}(1993)}]{griffin/orton:1993}
{Griffin}, M.~J., \& {Orton}, G.~S. 1993, Icarus, 105, 537

\bibitem[{{Halverson} {et~al.}(2008){Halverson}, {Lanting}, {Ade}, {Basu},
  {Bender}, {Benson}, {Bertoldi}, {Cho}, {Chon}, {Clarke}, {Dobbs}, {Ferrusca},
  {Guesten}, {Holzapfel}, {Kovacs}, {Kennedy}, {Kermish}, {Kneissl}, {Lee},
  {Lueker}, {Mehl}, {Menten}, {Muders}, {Nord}, {Pacaud}, {Plagge},
  {Reichardt}, {Richards}, {Schaaf}, {Schilke}, {Schuller}, {Schwan},
  {Spieler}, {Tucker}, {Weiss}, \& {Zahn}}]{halverson/etal:2008}
{Halverson}, N.~W. {et~al.} 2008, arXiv:0807:4208, 0807.4208

\bibitem[{Hincks {et~al.}(2008)Hincks, Ade, Allen, Amiri, Appel, Battistelli,
  Burger, Chervenak, Dahlen, Denny, Devlin, Dicker, Doriese, D{\"{u}}nner,
  Essinger-Hileman, Fisher, Fowler, Halpern, Hargrave, Hasselfield, Hilton,
  Irwin, Jarosik, Kaul, Klein, Lau, Limon, Lupton, Marriage, Martocci,
  Mauskopf, Moseley, Netterfield, Niemack, Nolta, Page, Parker, Sederberg,
  Staggs, Stryzak, Swetz, Switzer, Thornton, Tucker, Wollack, \&
  Zhao}]{hincks/etal:2008}
Hincks, A.~D. {et~al.} 2008, in Proc.\ SPIE, ed. W.~D. Duncan, W.~S. Holland,
  S.~Withington, \& J.~Zmuidzinas, Vol. 7020 (SPIE), 70201P

\bibitem[{{Hinshaw} {et~al.}(2007){Hinshaw}, {Nolta}, {Bennett}, {Bean},
  {Dor{\'e}}, {Greason}, {Halpern}, {Hill}, {Jarosik}, {Kogut}, {Komatsu},
  {Limon}, {Odegard}, {Meyer}, {Page}, {Peiris}, {Spergel}, {Tucker}, {Verde},
  {Weiland}, {Wollack}, \& {Wright}}]{hinshaw/etal:2007}
{Hinshaw}, G. {et~al.} 2007, \apjs, 170, 288, arXiv:astro-ph/0603451

\bibitem[{{Hughes} {et~al.}(2009){Hughes}, {Menanteau}, {Sehgal}, {Infante}, \&
  {Barrientos}}]{hughes/etal:2009}
{Hughes}, J.~P., {Menanteau}, F., {Sehgal}, N., {Infante}, L., \& {Barrientos},
  F. 2009, in Bulletin of the American Astronomical Society, Vol.~41, Bulletin
  of the American Astronomical Society, 336

\bibitem[{{Keih{\"a}nen} {et~al.}(2005){Keih{\"a}nen}, {Kurki-Suonio}, \&
  {Poutanen}}]{keihanen/kurki-suonio/poutanen:2005}
{Keih{\"a}nen}, E., {Kurki-Suonio}, H., \& {Poutanen}, T. 2005, \mnras, 360,
  390, arXiv:astro-ph/0412517

\bibitem[{{Keih{\"a}nen} {et~al.}(2004){Keih{\"a}nen}, {Kurki-Suonio},
  {Poutanen}, {Maino}, \& {Burigana}}]{keihanen/etal:2004}
{Keih{\"a}nen}, E., {Kurki-Suonio}, H., {Poutanen}, T., {Maino}, D., \&
  {Burigana}, C. 2004, \aap, 428, 287, arXiv:astro-ph/0304411

\bibitem[{{Komatsu} \& {Seljak}(2001)}]{komatsu/seljak:2001}
{Komatsu}, E., \& {Seljak}, U. 2001, \mnras, 327, 1353

\bibitem[{{Kramer} {et~al.}(2008){Kramer}, {Moreno}, \&
  {Greve}}]{kramer/moreno/greve:2008}
{Kramer}, C., {Moreno}, R., \& {Greve}, A. 2008, \aap, 482, 359, 0801.4452

\bibitem[{{Kuo} {et~al.}(2004){Kuo}, {Ade}, {Bock}, {Cantalupo}, {Daub},
  {Goldstein}, {Holzapfel}, {Lange}, {Lueker}, {Newcomb}, {Peterson}, {Ruhl},
  {Runyan}, \& {Torbet}}]{kuo/etal:2004}
{Kuo}, C.~L. {et~al.} 2004, \apj, 600, 32, arXiv:astro-ph/0212289

\bibitem[{{Maino} {et~al.}(2002){Maino}, {Burigana}, {G{\'o}rski}, {Mandolesi},
  \& {Bersanelli}}]{maino/etal:2002}
{Maino}, D., {Burigana}, C., {G{\'o}rski}, K.~M., {Mandolesi}, N., \&
  {Bersanelli}, M. 2002, \aap, 387, 356, arXiv:astro-ph/0202271

\bibitem[{{Markevitch} {et~al.}(2002){Markevitch}, {Gonzalez}, {David},
  {Vikhlinin}, {Murray}, {Forman}, {Jones}, \& {Tucker}}]{markevitch/etal:2002}
{Markevitch}, M., {Gonzalez}, A.~H., {David}, L., {Vikhlinin}, A., {Murray},
  S., {Forman}, W., {Jones}, C., \& {Tucker}, W. 2002, \apjl, 567, L27,
  arXiv:astro-ph/0110468

\bibitem[{{Marriage} {et~al.}(2010){Marriage}, {Acquaviva}, {Ade}, {Aguirre},
  {Amiri}, {Appel}, {Barrientos}, {Battistelli}, {Bond}, {Brown}, {Burger},
  {Chervenak}, {Das}, {Devlin}, {Dicker}, {Doriese}, {Dunkley}, {Dunner},
  {Essinger-Hileman}, {Fisher}, {Fowler}, {Hajian}, {Halpern}, {Hasselfield},
  {Hern'andez-Monteagudo}, {Hilton}, {Hilton}, {Hincks}, {Hlozek},
  {Huffenberger}, {Hughes}, {Hughes}, {Infante}, {Irwin}, {Juin}, {Kaul},
  {Klein}, {Kosowsky}, {Lau}, {Limon}, {Lin}, {Lupton}, {Marsden}, {Martocci},
  {Mauskopf}, {Menanteau}, {Moodley}, {Moseley}, {Netterfield}, {Niemack},
  {Nolta}, {Page}, {Parker}, {Partridge}, {Quintana}, {Reese}, {Reid},
  {Sehgal}, {Sherwin}, {Sievers}, {Spergel}, {Staggs}, {Swetz}, {Switzer},
  {Thornton}, {Trac}, {Tucker}, {Warne}, {Wilson}, {Wollack}, \&
  {Zhao}}]{marriage/etal:prepb}
{Marriage}, T.~A. {et~al.} 2010, ArXiv e-prints, 1010.1065

\bibitem[{Marriage {et~al.}(2006)Marriage, Chervenak, \&
  Doriese}]{marriage/chervenak/doriese:2006}
Marriage, T.~A., Chervenak, J.~A., \& Doriese, W.~B. 2006, Nuc Inst \& Meth. in
  Phys Res A, 559, 551

\bibitem[{{Marten} {et~al.}(2005){Marten}, {Matthews}, {Owen}, {Moreno},
  {Hidayat}, \& {Biraud}}]{marten/etal:2005}
{Marten}, A., {Matthews}, H.~E., {Owen}, T., {Moreno}, R., {Hidayat}, T., \&
  {Biraud}, Y. 2005, \aap, 429, 1097

\bibitem[{{Menanteau} {et~al.}(2010){Menanteau}, {Gonz{\'a}lez}, {Juin},
  {Marriage}, {Reese}, {Acquaviva}, {Aguirre}, {Appel}, {Baker}, {Barrientos},
  {Battistelli}, {Bond}, {Das}, {Deshpande}, {Devlin}, {Dicker}, {Dunkley},
  {D{\"u}nner}, {Essinger-Hileman}, {Fowler}, {Hajian}, {Halpern},
  {Hasselfield}, {Hern{\'a}ndez-Monteagudo}, {Hilton}, {Hincks}, {Hlozek},
  {Huffenberger}, {Hughes}, {Infante}, {Irwin}, {Klein}, {Kosowsky}, {Lin},
  {Marsden}, {Moodley}, {Niemack}, {Nolta}, {Page}, {Parker}, {Partridge},
  {Sehgal}, {Sievers}, {Spergel}, {Staggs}, {Swetz}, {Switzer}, {Thornton},
  {Trac}, {Warne}, \& {Wollack}}]{menanteau/etal:2010}
{Menanteau}, F. {et~al.} 2010, \apj, 723, 1523, 1006.5126

\bibitem[{{Menanteau} \& {Hughes}(2009)}]{menanteau/hughes:2009}
{Menanteau}, F., \& {Hughes}, J.~P. 2009, \apjl, 694, L136, 0811.3596

\bibitem[{{Meyer} {et~al.}(1991){Meyer}, {Cheng}, \&
  {Page}}]{meyer/cheng/page:1991}
{Meyer}, S.~S., {Cheng}, E.~S., \& {Page}, L.~A. 1991, \apjl, 371, L7

\bibitem[{{Niemack}(2006)}]{niemack:2006}
{Niemack}, M.~D. 2006, in Presented at the Society of Photo-Optical
  Instrumentation Engineers (SPIE) Conference, Vol. 6275, Society of
  Photo-Optical Instrumentation Engineers (SPIE) Conference Series, 62750C

\bibitem[{Niemack {et~al.}(2008)Niemack, Zhao, Wollack, Thornton, Switzer,
  Swetz, Staggs, Page, Stryzak, Moseley, Marriage, Limon, Lau, Klein, Kaul,
  Jarosik, Irwin, Hincks, Hilton, Halpern, Fowler, Fisher, D{\"{u}}nner,
  Doriese, Dicker, Devlin, Chervenak, Burger, Battistelli, Appel, Amiri, Allen,
  \& Aboobaker}]{niemack/etal:2008}
Niemack, M.~D. {et~al.} 2008, J. Low Temp. Phys., 151, 690

\bibitem[{{Nozawa} {et~al.}(2006){Nozawa}, {Itoh}, {Suda}, \&
  {Ohhata}}]{nozawa/etal:2006}
{Nozawa}, S., {Itoh}, N., {Suda}, Y., \& {Ohhata}, Y. 2006, Nuovo Cimento B
  Serie, 121, 487, arXiv:astro-ph/0507466

\bibitem[{{Page} {et~al.}(2003){Page}, {Barnes}, {Hinshaw}, {Spergel},
  {Weiland}, {Wollack}, {Bennett}, {Halpern}, {Jarosik}, {Kogut}, {Limon},
  {Meyer}, {Tucker}, \& {Wright}}]{page/etal:2003}
{Page}, L. {et~al.} 2003, \apjs, 148, 39, arXiv:astro-ph/0302214

\bibitem[{{P\'{e}rez-Beaupuits} {et~al.}(2005){P\'{e}rez-Beaupuits}, {Rivera},
  \& {Nyman}}]{perez-beaupuits/rivera/nyman:2005}
{P\'{e}rez-Beaupuits}, J.~P., {Rivera}, R.~C., \& {Nyman}, L.-A. 2005, Height
  and Velocity of the Turbulence Layer at Chajnantor Estimated From Radiometric
  Measurements, Memo 542, Atacama Large Millimeter Array (ALMA,
  Charlottesville, VA: NRAO)

\bibitem[{Press {et~al.}(1992)Press, Teukolsky, Vetterling, \&
  Flannery}]{press/teukolsky/vetterling:NRC:2e}
Press, W.~H., Teukolsky, S.~A., Vetterling, W.~T., \& Flannery, B.~P. 1992,
  {Numerical Recipes in C: The Art of Scientific Computing}, {2nd} edn.
  ({Cambridge University Press})

\bibitem[{{Reichardt} {et~al.}(2009{\natexlab{a}}){Reichardt}, {Ade}, {Bock},
  {Bond}, {Brevik}, {Contaldi}, {Daub}, {Dempsey}, {Goldstein}, {Holzapfel},
  {Kuo}, {Lange}, {Lueker}, {Newcomb}, {Peterson}, {Ruhl}, {Runyan}, \&
  {Staniszewski}}]{reichardt/etal:2009}
{Reichardt}, C.~L. {et~al.} 2009{\natexlab{a}}, \apj, 694, 1200, 0801.1491

\bibitem[{{Reichardt} {et~al.}(2009{\natexlab{b}}){Reichardt}, {Zahn}, {Ade},
  {Basu}, {Bender}, {Bertoldi}, {Cho}, {Chon}, {Dobbs}, {Ferrusca},
  {Halverson}, {Holzapfel}, {Horellou}, {Johansson}, {Johnson}, {Kennedy},
  {Kneissl}, {Lanting}, {Lee}, {Lueker}, {Mehl}, {Menten}, {Nord}, {Pacaud},
  {Richards}, {Schaaf}, {Schwan}, {Spieler}, {Weiss}, \&
  {Westbrook}}]{reichardt/etal:2009a}
------. 2009{\natexlab{b}}, \apj, 701, 1958, 0904.3939

\bibitem[{{Reiprich} \& {B{\"o}hringer}(2002)}]{reiprich/bohringer:2002}
{Reiprich}, T.~H., \& {B{\"o}hringer}, H. 2002, \apj, 567, 716,
  arXiv:astro-ph/0111285

\bibitem[{{Rephaeli}(1995)}]{rephaeli:1995}
{Rephaeli}, Y. 1995, \apj, 445, 33

\bibitem[{{Rose} {et~al.}(2002){Rose}, {Gaba}, {Christiansen}, {Davis},
  {Caldwell}, {Hunstead}, \& {Johnston-Hollitt}}]{rose/etal:2002}
{Rose}, J.~A., {Gaba}, A.~E., {Christiansen}, W.~A., {Davis}, D.~S.,
  {Caldwell}, N., {Hunstead}, R.~W., \& {Johnston-Hollitt}, M. 2002, \aj, 123,
  1216, arXiv:astro-ph/0112346

\bibitem[{Schroeder(2000)}]{schroeder:AO:2e}
Schroeder, D.~J. 2000, {Astronomical Optics}, {2nd} edn. ({Academic Press})

\bibitem[{{Schumaker}(2007)}]{schumaker:SFBT:3e}
{Schumaker}, L.~L. 2007, {Spline Functions: Basic Theory}, {3rd} edn.
  ({Cambridge University Press})

\bibitem[{{Sharp} {et~al.}(2010){Sharp}, {Marrone}, {Carlstrom}, {Culverhouse},
  {Greer}, {Hawkins}, {Hennessy}, {Joy}, {Lamb}, {Leitch}, {Loh}, {Miller},
  {Mroczkowski}, {Muchovej}, {Pryke}, \& {Woody}}]{sharp/etal:2010}
{Sharp}, M.~K. {et~al.} 2010, \apj, 713, 82, 0901.4342

\bibitem[{{Sievers} {et~al.}(2009){Sievers}, {Mason}, {Weintraub}, {Achermann},
  {Altamirano}, {Bond}, {Bronfman}, {Bustos}, {Contaldi}, {Dickinson}, {Jones},
  {May}, {Myers}, {Oyarce}, {Padin}, {Pearson}, {Pospieszalski}, {Readhead},
  {Reeves}, {Shepherd}, {Taylor}, \& {Torres}}]{sievers/etal:prep}
{Sievers}, J.~L. {et~al.} 2009, arXiv:0901.4540, 0901.4540

\bibitem[{{Staniszewski} {et~al.}(2009){Staniszewski}, {Ade}, {Aird}, {Benson},
  {Bleem}, {Carlstrom}, {Chang}, {Cho}, {Crawford}, {Crites}, {de Haan},
  {Dobbs}, {Halverson}, {Holder}, {Holzapfel}, {Hrubes}, {Joy}, {Keisler},
  {Lanting}, {Lee}, {Leitch}, {Loehr}, {Lueker}, {McMahon}, {Mehl}, {Meyer},
  {Mohr}, {Montroy}, {Ngeow}, {Padin}, {Plagge}, {Pryke}, {Reichardt}, {Ruhl},
  {Schaffer}, {Shaw}, {Shirokoff}, {Spieler}, {Stalder}, {Stark},
  {Vanderlinde}, {Vieira}, {Zahn}, \& {Zenteno}}]{staniszewski/etal:2009}
{Staniszewski}, Z. {et~al.} 2009, \apj, 701, 32, 0810.1578

\bibitem[{{Sunyaev} \& {Zel'dovich}(1970)}]{sunyaev/zeldovich:1970}
{Sunyaev}, R.~A., \& {Zel'dovich}, Y.~B. 1970, Comments on Astrophysics and
  Space Physics, 2, 66

\bibitem[{{Sutton} {et~al.}(2009){Sutton}, {Johnson}, {Brown}, {Cabella},
  {Ferreira}, \& {Smith}}]{sutton/etal:2009}
{Sutton}, D., {Johnson}, B.~R., {Brown}, M.~L., {Cabella}, P., {Ferreira},
  P.~G., \& {Smith}, K.~M. 2009, \mnras, 393, 894, 0807.3658

\bibitem[{Swetz {et~al.}(2008)Swetz, Ade, Allen, Amiri, Appel, Battistelli,
  Burger, Chervenak, Dahlen, Das, Denny, Devlin, Dicker, Doriese, D{\"{u}}nner,
  Essinger-Hileman, Fisher, Fowler, Gao, Hajian, Halpern, Hargrave,
  Hasselfield, Hilton, Hincks, Irwin, Jarosik, Kaul, Klein, Knotek, Lau, Limon,
  Lupton, Marriage, Martocci, Mauskopf, Moseley, Netterfield, Niemack, Nolta,
  Page, Parker, Reid, Reintsema, Sederberg, Sehgal, Sievers, Spergel, Staggs,
  Stryzak, Switzer, Thornton, Tucker, Wollack, \& Zhao}]{swetz/etal:2008}
Swetz, D.~S. {et~al.} 2008, in Proc.\ SPIE, ed. W.~D. Duncan, W.~S. Holland,
  S.~Withington, \& J.~Zmuidzinas, Vol. 7020 (SPIE), 702008

\bibitem[{Switzer {et~al.}(2008)Switzer, Allen, Amiri, Appel, Battistelli,
  Burger, Chervenak, Dahlen, Das, Devlin, Dicker, Doriese, D\"{u}nner,
  Essinger-Hileman, Gao, Halpern, Hasselfield, Hilton, Hincks, Irwin, Knotek,
  Fisher, Fowler, Jarosik, Kaul, Klein, Lau, Limon, Lupton, Marriage, Martocci,
  Moseley, Netterfield, Niemack, Nolta, Page, Parker, Reid, Reintsema,
  Sederberg, Sievers, Spergel, Staggs, Stryzak, Swetz, Thornton, Wollack, \&
  Zhao}]{switzer/etal:2008}
Switzer, E.~R. {et~al.} 2008, in Proc.\ SPIE, ed. A.~Bridger \& N.~M.
  Radziwill, Vol. 7019 (SPIE), 70192L

\bibitem[{{Tegmark}(1997)}]{tegmark:1997a}
{Tegmark}, M. 1997, \apjl, 480, L87, arXiv:astro-ph/9611130

\bibitem[{Thornton {et~al.}(2008)Thornton, Ade, Allen, Amiri, Appel,
  Battistelli, Burger, Chervenak, Devlin, Dicker, Doriese, Essinger-Hileman,
  Fisher, Fowler, Halpern, Hargrave, Hasselfield, Hilton, Hincks, Irwin,
  Jarosik, Kaul, Klein, Lau, Limon, Marriage, Martocci, Mauskopf, Moseley,
  Niemack, Page, Parker, Reidel, Reintsema, Staggs, Stryzak, Swetz, Switzer,
  Tucker, Wollack, \& Zhao}]{thornton/etal:2008}
Thornton, R.~J. {et~al.} 2008, in Proc.\ SPIE, ed. W.~D. Duncan, W.~S. Holland,
  S.~Withington, \& J.~Zmuidzinas, Vol. 7020 (SPIE), 70201R

\bibitem[{{Tucker} {et~al.}(1998){Tucker}, {Blanco}, {Rappoport}, {David},
  {Fabricant}, {Falco}, {Forman}, {Dressler}, \& {Ramella}}]{tucker/etal:1998}
{Tucker}, W. {et~al.} 1998, \apjl, 496, L5, arXiv:astro-ph/9801120

\bibitem[{{Umetsu} {et~al.}(2009){Umetsu}, {Birkinshaw}, {Liu}, {Wu},
  {Medezinski}, {Broadhurst}, {Lemze}, {Zitrin}, {Ho}, {Huang}, {Koch}, {Liao},
  {Lin}, {Molnar}, {Nishioka}, {Wang}, {Altamirano}, {Chang}, {Chang}, {Chang},
  {Chen}, {Han}, {Huang}, {Hwang}, {Jiang}, {Kesteven}, {Kubo}, {Li},
  {Martin-Cocher}, {Oshiro}, {Raffin}, {Wei}, \& {Wilson}}]{umetsu/etal:2009}
{Umetsu}, K. {et~al.} 2009, \apj, 694, 1643, 0810.0969

\bibitem[{{Verde} {et~al.}(2002){Verde}, {Haiman}, \&
  {Spergel}}]{verde/haiman/spergel:2002}
{Verde}, L., {Haiman}, Z., \& {Spergel}, D.~N. 2002, \apj, 581, 5,
  arXiv:astro-ph/0106315

\bibitem[{{Voges} {et~al.}(1999){Voges}, {Aschenbach}, {Boller},
  {Br{\"a}uninger}, {Briel}, {Burkert}, {Dennerl}, {Englhauser}, {Gruber},
  {Haberl}, {Hartner}, {Hasinger}, {K{\"u}rster}, {Pfeffermann}, {Pietsch},
  {Predehl}, {Rosso}, {Schmitt}, {Tr{\"u}mper}, \&
  {Zimmermann}}]{voges/etal:1999}
{Voges}, W. {et~al.} 1999, \aap, 349, 389, arXiv:astro-ph/9909315

\bibitem[{{Werner} {et~al.}(2007){Werner}, {Churazov}, {Finoguenov},
  {Markevitch}, {Burenin}, {Kaastra}, \& {B{\"o}hringer}}]{werner/etal:2007}
{Werner}, N., {Churazov}, E., {Finoguenov}, A., {Markevitch}, M., {Burenin},
  R., {Kaastra}, J.~S., \& {B{\"o}hringer}, H. 2007, \aap, 474, 707, 0708.3253

\bibitem[{{White} \& {Srednicki}(1995)}]{white/srednicki:1995}
{White}, M., \& {Srednicki}, M. 1995, \apj, 443, 6, arXiv:astro-ph/9402037

\bibitem[{{Zeldovich} \& {Sunyaev}(1969)}]{zeldovich/sunyaev:1969}
{Zeldovich}, Y.~B., \& {Sunyaev}, R.~A. 1969, \apss, 4, 301

\bibitem[{{Zhang} {et~al.}(2006){Zhang}, {B{\"o}hringer}, {Finoguenov},
  {Ikebe}, {Matsushita}, {Schuecker}, {Guzzo}, \& {Collins}}]{zhang/etal:2006}
{Zhang}, Y.-Y., {B{\"o}hringer}, H., {Finoguenov}, A., {Ikebe}, Y.,
  {Matsushita}, K., {Schuecker}, P., {Guzzo}, L., \& {Collins}, C.~A. 2006,
  \aap, 456, 55, arXiv:astro-ph/0603275

\bibitem[{{Zhang} {et~al.}(2008){Zhang}, {Finoguenov}, {B{\"o}hringer},
  {Kneib}, {Smith}, {Kneissl}, {Okabe}, \& {Dahle}}]{zhang/etal:2008}
{Zhang}, Y.-Y., {Finoguenov}, A., {B{\"o}hringer}, H., {Kneib}, J.-P., {Smith},
  G.~P., {Kneissl}, R., {Okabe}, N., \& {Dahle}, H. 2008, \aap, 482, 451,
  0802.0770

\bibitem[{{Zhao} {et~al.}(2008){Zhao}, {Allen}, {Amiri}, {Appel},
  {Battistelli}, {Burger}, {Chervenak}, {Dahlen}, {Denny}, {Devlin}, {Dicker},
  {Doriese}, {D{\"u}nner}, {Essinger-Hileman}, {Fisher}, {Fowler}, {Halpern},
  {Hilton}, {Hincks}, {Irwin}, {Jarosik}, {Klein}, {Lau}, {Marriage},
  {Martocci}, {Moseley}, {Niemack}, {Page}, {Parker}, {Sederberg}, {Staggs},
  {Stryzak}, {Swetz}, {Switzer}, {Thornton}, \& {Wollack}}]{zhao/etal:2008}
{Zhao}, Y. {et~al.} 2008, in Proc.\ SPIE, ed. W.~D. Duncan, W.~S. Holland,
  S.~Withington, \& J.~Zmuidzinas, Vol. 7020 (SPIE), 70200O

\end{thebibliography}
\bibliographystyle{hapj}

\end{document}